\documentclass[aps,preprint]{revtex4-1}
\usepackage{amsmath}
\usepackage{graphicx}
\usepackage{color}

\begin{document}

\title{Density Functional Theory of Inhomogeneous Liquids IV.
Squared-gradient approximation and Classical Nucleation Theory}
\author{James F. Lutsko}
\affiliation{Center for Nonlinear Phenomena and Complex Systems CP 231, Universit\'{e} Libre de Bruxelles, Blvd. du
Triomphe, 1050 Brussels, Belgium}
\email{jlutsko@ulb.ac.be}
\homepage{http://www.lutsko.com}

\begin{abstract}
The Squared-Gradient approximation to the Modified-Core Van der Waals
density functional theory model is developed. A simple, explicit expression
for the SGA coefficient involving only the bulk equation of state and the
interaction potential is given. The model is solved for planar interfaces
and spherical clusters and is shown to be quantitatively accurate in
comparisons to computer simulations. An approximate technique for solving
the SGA based on piecewise-linear density profiles is introduced and is
shown to give reasonable zeroth-order approximations to the numerical
solution of the model. The piecewise-linear models of spherical clusters are
shown to be a natural extension of Classical Nucleation Theory and serve to
clarify some of the non-classical effects previously observed in
liquid-vapor nucleation. Nucleation pathways are investigated using both
constrained energy-minimization and steepest-descent techniques.
\end{abstract}

\date{\today }
\maketitle

\section{Introduction}

The description of inhomogeneous systems remains one of the most important
problems in several areas of physics. Recently, it has been shown that
classical Density\ Functional Theory (DFT)\ can give quantitatively accurate
results for many inhomogeneous systems including the structure and surface
tension of the liquid-vapor interface, confined fluids near walls, in slit
pores\cite{Lutsko_JCP_2008} and in small cavities\cite{0953-8984-22-3-035101}
and of liquid-vapor nucleation\cite{Lutsko_JCP_2008_2}. While these results
confirm that DFT can be a useful tool for making quantitatively accurate
calculations, they provide limited physical insight due to the complexity of
the DFT models and the need for extensive numerical calculations. 

One simplification of DFT is to relate it to more physically-inspired
descriptions of inhomogeneous systems. The oldest, and in many contexts most
important, such model is the squared-gradient approximation (SGA)\ that
dates back to van der Waals\cite{VDW1,VDW2} and has been re-invented and
exploited in many different circumstances, most notably by Landau in the
context of phase transitions\cite{GL} and by Cahn and Hilliard\cite%
{CahnHilliard} for the description of interfaces. Evans established the link
whereby SGA is an approximation to more fundamental DFT\cite{Evans1979} and
this link has more recently been refined and systematized\cite%
{Lowen1,Lowen2,LutskoGL}. Despite this justification for SGA, it
nevertheless has the reputation of giving only a qualitative, at best
semi-quantitative, approximation to real systems\cite{Cornelisse}. One
result reported here is that SGA based on DFT can in fact be surprisingly
accurate.

If DFT can be well approximated by SGA, then a substantial reduction of
complexity has been achieved. However, for most applications, SGA must also
be solved numerically. No matter what details go into the models, this
fundamentally involves solving for the spatially-varying density which
minimizes the free energy. A second development presented here is the use of
piecewise-linear approximations for the density profiles. In the simplest
case, this involves approximating the interface between two bulk phases as a
linear function. The calculation can be systematically improved by replacing
the single linear function by a sequence of linear functions, or links, so
that the true profile is obtained in the limit of the number of links going
to infinite. It is shown that the simplest single-link profile provides a
rather accurate approximation to the numerical solution of the SGA with the
benefit of giving nearly analytic expressions for surface tensions and
interfacial width of planar interfaces and of interfacial width, surface
tension and Tolman length for spherical clusters. An additional benefit is
that by reducing the DFT integral theory to a simple algebraic model, it
gives a systematic link between DFT and Classical Nucleation Theory (CNT).

Finally, the algebraic model for spherical clusters is used to study the
nucleation pathway for homogeneous liquid-vapor nucleation. A first method
is to minimize the free energy subject to constraints such as fixed number
of atoms in the cluster or fixed radius of the cluster. It is found these
and other constraints are all problematic and fail to give a consistent
picture of homogeneous nucleation. An alternative method is the construction
of steepest-descent paths in density space linking the transition state
(i.e. the critical cluster) to the bulk liquid and bulk vapor free energy
minima\cite{Wales}. This is found to give a simple and plausible description
of the nucleation pathway.

In the next Section, the SGA is formulated based on the quantitatively
accurate Modified-Core Van der Waals model DFT. The result is an SGA model
for fluids that requires only the bulk equation of state and the interaction
potential as input. To the extent that the equation of state can be
accurately approximation by, say, thermodynamic perturbation theory, only
the interaction potential need be specified. The application to planar
interfaces and for spherical clusters and the piecewise-linear approximation
are also discussed. In the third Section, this model is used to explore
liquid-vapor nucleation. First, the connection with the ideas that underlie
Classical Nucleation Theory is made. There, different methods of formulating
the problem of the determination of the nucleation pathway are described and
their relative merits compared. The fourth Section gives a comparison of the
SGA and the analytic models to computer simulation of both planar interfaces
and clusters. The paper ends with a discussion of the results.

\section{Theory}

\subsection{Squared-gradient approximation}

Density Functional Theory is an approach to equilibrium
statistical mechanics which is formulated in the grand-canonical ensemble at
constant temperature, $T$, chemical potential, $\mu $, and volume $V$. The
fundamental object is the local number density $\rho \left( \mathbf{r}%
\right) $\cite{HansenMcdonald, Evans1979, Evans1992}. It can be shown that
there exists a functional of the local density, $\Omega \left[ \rho \right] $%
, having the property that it is minimized by the equilibrium density
function and that its value at this minimum is the grand-potential for the
system. It can be written as%
\begin{equation}
\Omega \left[ \rho \right] =F\left[ \rho \right] +\int \left( \phi \left( 
\mathbf{r}\right) -\mu \right) \rho \left( \mathbf{r}\right) d\mathbf{r}
\end{equation}%
where $\phi \left( \mathbf{r}\right) $ is any external field that may act on
the system and where $F\left[ \rho \right] $ is independent of the field but
otherwise unknown except for special cases such as the ideal gas for which 
\begin{equation}
F_{id}\left[ \rho \right] =k_{B}T\int \rho \left( \mathbf{r}\right) \left(
\ln \rho \left( \mathbf{r}\right) -1\right) d\mathbf{r}
\end{equation}%
where $k_{B}$ is Boltzmann's constant. When evaluated at the equilibrium
density, $F\left[ \rho \right] $ is the intrinsic Helmholtz free energy. Minimization
of $\Omega \left[ \rho \right] $ gives the Euler-Lagrange equation%
\begin{equation}
\frac{\delta F\left[ \rho \right] }{\delta \rho \left( \mathbf{r}\right) }%
+\phi \left( \mathbf{r}\right) -\mu =0  \label{EL}
\end{equation}%
A key result used to guide the formation of models is that the excess
functional $F_{ex}\left[ \rho \right] =F\left[ \rho \right] -F_{id}\left[
\rho \right] $ is related to the two-body direct correlation function via%
\begin{equation}
c\left( \mathbf{r}_{1},\mathbf{r}_{2};\left[ \rho \right] \right) =-\frac{%
\delta ^{2}\beta F_{ex}\left[ \rho \right] }{\delta \rho \left( \mathbf{r}%
_{1}\right) \delta \rho \left( \mathbf{r}_{2}\right) }
\end{equation}%
where $\beta =1/k_{B}T$. By definition, a bulk fluid is a system with
constant equilibrium density, $\rho \left( \mathbf{r}\right) =\overline{\rho 
}$ in which case $F\left[ \rho \right] =F\left( \overline{\rho }\right) $ is
the usual Helmholtz free energy function for the fluid.

It can be shown\cite{LutskoGL} that a systematic expansion of $F\left[ \rho %
\right] $ takes the form%
\begin{equation}
F\left[ \rho \right] =\int_{V}\left\{ f\left( \rho \left( \mathbf{r}\right)
\right) +\frac{1}{2}K\left( \rho \left( \mathbf{r}\right) \right) \left( 
\mathbf{\nabla }\rho \left( \mathbf{r}\right) \right) ^{2}+...\right\} d%
\mathbf{r}  \label{SGA}
\end{equation}%
where $f\left( \overline{\rho }\right) =\frac{1}{V}F\left( \overline{\rho }%
\right) $ is the Helmholtz free energy per unit volume of the homogeneous
system, the ellipses indicate higher order terms in the gradients and the
coefficient of the second order term is 
\begin{equation}
\beta K\left(\rho(\mathbf{r})\right) =\frac{1}{6V}\int c\left( \mathbf{r}%
_{12};\rho (\mathbf{r})\right) r_{12}^{2}d\mathbf{r}_{1}d\mathbf{r}_{2}
\end{equation}%
where $c\left( \mathbf{r}_{12};\overline{\rho }\right) =c\left( \mathbf{r}%
_{1},\mathbf{r}_{2};\overline{\rho }\right) $ is the translationally
invariant direct correlation function of the bulk state. The gradient
expansion is thus seen as a link between the properties of the bulk state -
which are generally accessible - and those of the inhomogeneous state which
are generally much more difficult to determine. Truncating the expansion at
second order gives the Squared-Gradient Approximation (SGA). Substituting
into Eq.(\ref{EL}) and taking the external potential to be zero gives%
\begin{equation}
\mathbf{\nabla }\cdot K\left( \rho \left( \mathbf{r}\right) \right) \mathbf{%
\nabla }\rho \left( \mathbf{r}\right) -\frac{1}{2}\left( \frac{\partial }{%
\partial \rho \left( \mathbf{r}\right) }K\left( \rho \left( \mathbf{r}%
\right) \right) \right) \left( \mathbf{\nabla }\rho \left( \mathbf{r}\right)
\right) ^{2}-\frac{\partial \omega \left( \rho \left( \mathbf{r}\right)
\right) }{\partial \rho \left( \mathbf{r}\right) }=0  \label{SGA-1}
\end{equation}%
where $\omega \left( \rho \right) =f\left( \rho \right) -\mu \rho $ is the
grand potential per unit volume. 

To implement the theory, two elements are necessary: the Helmholtz free
energy and the direct correlation function of the bulk system. These are not
independent: the equation of state is easily calculated from the DCF via the
compressibility equation as discussed below. For realistic potentials, the
DCF can be calculated using liquid state theory such as the Percus-Yevick or
the HNC approximations. However, when dealing with an interfacial system,
the density typically varies a lot:\ for a liquid-vapor interface it
obviously spans the range from (dense) liquid to (low density) vapor. In
particular, it passes through densities that lie in the two-phase region of
the bulk phase diagram and liquid-state theories often do not have solutions
in those regions. While methods of circumventing this problem by means of
e.g. interpolation have been proposed (see e.g. Ref.\cite{Cornelisse}) this
remains a problematic issue. Of course, this is an issue facing all DFT's
since the free energy functional is always related to the DCF so that it
might be suspected that a successful DCF\ would imply some means around this
problem. The Modified-Core Van der Waals model DFT is based on a simple
approximation to the DCF and gives good results for a wide variety of
interfacial systems\cite{Lutsko_JCP_2008,Lutsko_JCP_2008_2}. The idea behind
it is to begin with the simplest Van der Waals model whereby the DCF is
approximated as that of a hard-sphere system with a mean-field treatment of
the attractive tail of the interaction,%
\begin{equation}
c_{VDW}\left( \mathbf{r}_{12};\overline{\rho }\right) =c_{HS}\left( \mathbf{r%
}_{12};\overline{\rho },d\right) -\beta v\left( r_{12}\right) \Theta \left(
r-d\right) 
\end{equation}%
where $v(r)$ is the molecular pair interaction potential, $d$ is the effective hard-sphere diameter, $c_{HS}\left( \mathbf{r}%
_{1},\mathbf{r}_{2};\overline{\rho },d\right) $ is the hard-sphere DCF and $%
\Theta (x)=1$ for $x>0$ and zero otherwise. Because of the relation between
the DCF and the excess free energy of the uniform bulk fluid,%
\begin{equation}
\beta f_{ex}\left( \bar{\rho}\right) \equiv \beta f\left( \overline{\rho }%
\right) -\beta f_{id}\left( \overline{\rho }\right) =-\frac{1}{V}\int_{0}^{%
\overline{\rho }}d\rho _{2}\int_{0}^{\rho _{2}}d\rho _{1}\int \int c\left( 
\mathbf{r}_{12};\rho _{1}\right) d\mathbf{r}_{1}d\mathbf{r}_{2},  \label{EOS}
\end{equation}%
this implies a rather inaccurate equation of state. Furthermore, it is
discontinuous at the hard-sphere boundary. To address both of these
problems, a linear correction is added to the core region giving the full
approximation\cite{Lutsko_JCP_2007}%
\begin{equation}
c\left( \mathbf{r}_{12};\overline{\rho }\right) =c_{HS}\left( \mathbf{r}%
_{12};\overline{\rho },d\right) +\left( b_{0}+b_{1}\frac{r}{d}\right) \Theta
\left( d-r\right) -\beta v\left( r_{12}\right) \Theta \left( r-d\right) 
\end{equation}%
The coefficients $b_{0}$ and $b_{1}$ depend on both density and temperature
and are chosen to reproduce a given equation of state and to make the DCF
continuous. This has been shown to give a rather good approximation for
dense fluids\cite{Lutsko_JCP_2007}. At low density, one has the exact result%
\begin{equation}
\lim_{\overline{\rho }\rightarrow 0}c\left( \mathbf{r}_{12};\overline{\rho }%
\right) =-\left( 1-e^{-\beta v\left( r_{12}\right) }\right) 
\end{equation}%
and is clearly not reproduced by the simple model. This limit can be
enforced by generalizing to%
\begin{equation}
c\left( \mathbf{r}_{12};\overline{\rho }\right) =c_{HS}\left( \mathbf{r}%
_{12};\overline{\rho },d\right) +\left( c\left( \mathbf{r}_{12};0\right)
-c_{HS}\left( \mathbf{r}_{12};0\right) \right) +\left( b_{0}+b_{1}\frac{r}{d}%
\right) \Theta \left( d-r\right) 
\end{equation}%
which also gives a reasonable approximation. However, as most properties are
insensitive to the low-density limit of the DCF, it is more convenient and
not much less accurate to work with the simpler approximation given above.

Assuming the hard-core diameter is chosen to be independent of density, the
core-correction coefficients are fixed by the requirements that the model
reproduce the given bulk equation of state, $f\left( \bar{\rho}\right) $,
and that the DCF be continuous giving 
\begin{align}
\frac{\partial ^{2}}{\partial \overline{\rho }^{2}}f\left( \overline{\rho }%
\right) & =\frac{\partial ^{2}}{\partial \overline{\rho }^{2}}f_{HS}\left( 
\overline{\rho }\right) -4\pi d^{3}\left( \frac{1}{3}b_{0}+\frac{1}{4}%
b_{1}\right) +4\pi \int_{d}^{\infty }\beta v\left( r\right) r^{2}dr
\label{bb} \\
c_{HS}\left( d_{-};\overline{\rho },d\right) +b_{0}+b_{1}& =-\beta v\left(
d\right)   \notag
\end{align}%
As shown in Appendix \ref{AppA}, using common models for the hard-sphere
DCF, the final result for the SGA coefficient takes the simple form%
\begin{equation}
K=-\frac{\pi d^{5}}{180}a_{3}\left( \overline{\rho }d^{3}\right) -\frac{%
d^{2}}{15}\frac{\partial ^{2}}{\partial \overline{\rho }^{2}}\beta
f_{ex}\left( \overline{\rho }\right) +\frac{2\pi }{45}\int_{d}^{\infty
}\left( 3r^{5}-2d^{2}r^{3}\right) \frac{d\beta v\left( r\right) }{dr}dr 
\end{equation}
where $a_{3}\left( \overline{\rho }d^{3}\right) $ depends on the particular
hard-sphere model used (explicit expressions are given in Appendix \ref{AppA}%
). However, since the coefficient of $a_{3}\left( \overline{\rho }\right) $
is so small, this term contributes only about $1\%$ to the over-all value of 
$K$ and can usually be safely neglected. Thus, in this model, the SGA
coefficient is a relatively simple function of the hard-sphere diameter, the
potential and the equation of state. While any reasonable value for the
hard-sphere diameter could be used, the calculations presented below are
based on the Barker-Henderson formula\cite{BarkerHend,HansenMcdonald},%
\begin{equation}
d=\int_{0}^{r_{0}}\left( 1-\exp \left( -\beta v\left( r\right) \right)
\right) dr,  \label{BH-HSD}
\end{equation}%
where $r_{0}$ is the smallest solution of $v\left( r_{0}\right) =0$.

\subsection{Planar and spherical interfaces}

A stable planar interface can only exist when the liquid and vapor are at
conditions of coexistence so that the value of the chemical potential is $%
\mu _{coex}$ and the bulk liquid and vapor densities are $\rho _{l,coex}$
and $\rho _{v,coex}$ respectively. (All of these quantities are
temperature-dependent.) Then, one can impose a density profile that varies
in only one dimension, $\rho \left( z\right) $ so that Eq.(\ref{SGA-1})
becomes%
\begin{equation}
\frac{d}{dz}K\left( \rho \left( z\right) \right) \frac{d}{dz}\rho \left(
z\right) -\frac{1}{2}\left( \frac{d}{d\rho \left( z\right) }K\left( \rho
\left( z\right) \right) \right) \left( \frac{d}{dz}\rho \left( z\right)
\right) ^{2}-\frac{d\omega \left( \rho \left( z\right) \right) }{d\rho
\left( z\right) }=0
\end{equation}%
Assuming that the density takes the value of the bulk liquid and vapor far
from the boundary, and noting that $\omega \left( \rho _{l,coex}\right)
=\omega \left( \rho _{c,coex}\right) \equiv \omega _{coex}$ the profile
equation can be integrated to get 
\begin{equation}
K\left( \rho \left( z\right) \right) \left( \frac{\partial }{\partial z}\rho
\left( z\right) \right) ^{2}=2\left( \omega \left( \rho \left( z\right)
\right) -\omega _{coex}\right)  \label{ELP}
\end{equation}%
This allows the excess free energy, hereafter referred to as the "surface
tension", to be evaluated as 
\begin{equation}
\frac{\Omega -\Omega _{coex}}{A}=\int_{\rho _{v,coex}}^{\rho _{l,coex}}\sqrt{%
2\left( \omega \left( \rho \right) -\omega _{coex}\right) K\left( \rho
\right) }d\rho  \label{gplanar}
\end{equation}%
while the actual profile must be obtained by integrating Eq.(\ref{ELP})
numerically.

Away from coexistence, a planar interface is unstable. In this case, it is
more pertinent to study clusters since an unstable phase will transform to a
stable phase by the formation and subsequent growth of a critical cluster.
Assuming spherical symmetry, the equation for a (meta-)stable profile becomes%
\begin{equation} \label{SEL}
K\left( \rho \left( r\right) \right) \frac{1}{r}\frac{d^{2}}{dr^{2}}r\rho
\left( r\right) +\frac{dK\left( \rho \left( r\right) \right) }{dr}\frac{d}{dr%
}\rho \left( r\right) -\frac{1}{2}\left( \frac{d}{d\rho \left( r\right) }%
K\left( \rho \left( r\right) \right) \right) \left( \frac{d}{dr}\rho \left(
r\right) \right) ^{2}-\frac{d\omega \left( \rho \left( r\right) \right) }{%
d\rho \left( r\right) }=0
\end{equation}%
and this does not admit of an exact quadrature. The only nontrivial solution
will correspond to the critical cluster in which the density near the origin
will be close to that of the stable phase, while the bulk (i.e. the density
far from the origin) will be that of the unstable phase.

\subsection{Analytic approximations}

An alternative to solving these equations numerically is to assume some
ansatz for the density, $\rho \left( \mathbf{r}\right) =\rho\left( \mathbf{r}%
;\Gamma \right) $, where the quantity on the right has a specified spatial
dependence, e.g. a sigmoidal function for the case of a planar interface,
and where $\Gamma $ represents a collection of parameters (e.g. the width
and center of the sigmoidal function). In fact, this procedure was been used for the planar interface by Telo da Gama and Evans\cite{GamaEvans} and, recently, for droplets by Ghosh and Ghosh\cite{Ghosh}. Then, rather than solving Eq.(\ref%
{SGA-1}) to get the profile, one substitutes the ansatz into the expression
for $\Omega \left[ \rho \right] $ and extremizes the free energy with
respect to the parameters $\Gamma $. For most cases, this will still involve
numerical calculations however more progress is possible if one assumes the
simplest reasonable approximation which is a piecewise-continuous profile.
Thus, for the planar profile, one can try%
\begin{equation}
\rho \left( z\right) =\left\{ 
\begin{array}{c}
\rho _{-\infty },\;z<-\frac{w}{2} \\ 
\rho _{-\infty }+\left( \rho _{\infty }-\rho _{-\infty }\right) \frac{z+w/2}{%
w},\;-\frac{w}{2}<z<\frac{w}{2} \\ 
\rho _{\infty },\;\frac{w}{2}<z%
\end{array}%
\right.
\end{equation}%
There are three parameters characterizing this profile: the densities $\rho
_{\pm \infty }$ on either side of the interface and the width of the
interface. The profile is continuous and differentiable except at $z=\pm 
\frac{w}{2}$ but this is sufficient to allow evaluation of the free energy.
It is easy to see that the minimization of the free energy for the case that
the volume is very large requires that 
\begin{equation}
\frac{\partial f\left( \rho _{\infty }\right) }{\partial \rho _{\infty }}=%
\frac{\partial f\left( \rho _{-\infty }\right) }{\partial \rho _{-\infty }}%
=\mu
\end{equation}%
so that a non-trivial interface is only possible at coexistence in which
case one of the densities must be that of the coexisting liquid and the
other that of the coexisting vapor. Then, the excess free energy per unit area is 
\begin{equation}
\gamma =\frac{\Omega -\Omega _{coex}}{A}=\int_{-\frac{w}{2}}^{\frac{w}{2}%
}\left\{ \omega \left( \rho \left( z\right) \right) -\omega _{coex}+\frac{1}{%
2}K\left( \rho \left( z\right) \right) \left( \frac{\rho _{\infty }-\rho
_{-\infty }}{w}\right) ^{2}\right\} dz
\end{equation}%
or, more simply,%
\begin{equation}
\gamma =w\left( \overline{\omega } -\omega _{coex}\right) +\frac{1}{2}\frac{%
\left( \rho _{\infty }-\rho _{-\infty }\right) ^{2}}{w}\overline{K}
\end{equation}%
with%
\begin{eqnarray}
\overline{\omega } &=&\frac{1}{\rho _{\infty }-\rho _{-\infty }}\int_{\rho
_{-\infty }}^{\rho _{\infty }}\omega \left( \rho \right) d\rho \\
\overline{K} &=&\frac{1}{\rho _{\infty }-\rho _{-\infty }}\int_{\rho
_{-\infty }}^{\rho _{\infty }}K\left( \rho \right) d\rho  \notag
\end{eqnarray}%
(For simplicity of notation, the dependence of $\bar{\omega}$ and $\bar{K}$
on the densities is not indicated explicitly). Minimizing with respect to
the width gives%
\begin{equation}
w=\sqrt{\frac{\left( \rho _{\infty }-\rho _{-\infty }\right) ^{2}\overline{K}
}{2\left( \overline{\omega } -\omega _{coex}\right) }}
\end{equation}%
and%
\begin{equation}
\gamma =\sqrt{2\left( \rho _{\infty }-\rho _{-\infty }\right) ^{2}\left( 
\overline{\omega } -\omega _{coex}\right) \overline{K} }
\end{equation}%
which is very similar to the exact result given in Eq.(\ref{gplanar}).
However, in the present case, one has analytic expressions as well for the
entire profile, including the width which can be written as%
\begin{equation}
w=\frac{\gamma }{2\left( \overline{\omega } -\omega _{coex}\right) }=\frac{%
\left( \rho _{\infty }-\rho _{-\infty }\right) ^{2}\overline{K} }{\gamma }
\end{equation}%
giving a simple relation between the interfacial width, the SGA coefficient
and the surface tension. Notice that using the explicit expression for the
SGA coefficient and neglecting the small hard-sphere term, one has that%
\begin{equation}
\overline{K} =\frac{2\pi }{45}\int_{d}^{\infty }\left( 3r^{5}-2d^{2}r^{3}\right) \frac{%
d\beta v\left( r\right) }{dr}dr  
\end{equation}
since  the integral of the density-dependent part of the coefficient gives no contribution at
coexistence. In fact, if the density dependence of $K(\rho)$ is weak, as
will be seen to be the case for a simple fluid, then the approximation $%
K(\rho) \rightarrow \bar{K}$ should be adequate thus giving an even simpler expression for the gradient coefficient based solely on the interaction potential.

One can make a similar ansatz for the spherical cluster, 
\begin{equation}
\rho \left( r\right) =\left\{ 
\begin{array}{c}
\rho _{0},\;r<R \\ 
\rho _{0}+\left( \rho _{\infty }-\rho _{0}\right) \frac{r-R}{w},\;R<r<R+w \\ 
\rho _{\infty },\;R+w<r%
\end{array}%
\right.
\end{equation}%
so that, with $\Omega_{\infty} \equiv V \omega(\rho_{\infty})$ and $\Delta
\Omega \equiv \Omega - \Omega_{\infty}$, 
\begin{eqnarray}
\Delta \Omega &=&\frac{4\pi }{3}R^{3}\Delta \omega  \label{cluster} \\
&&+4\pi R^{2}w\left( \overline{\omega}_{0}+2\overline{\omega }_{1}\left( 
\frac{w}{R}\right) +\overline{\omega }_{2}\left( \frac{w}{R}\right)
^{2}+\left( \frac{\left( \rho _{\infty }-\rho _{0}\right) ^{2}}{2w^{2}}%
\right) \left( \overline{K}_{0}+2\overline{K}_{1}\left( \frac{w}{R}\right) +%
\overline{K}_{2}\left( \frac{w}{R}\right) ^{2}\right) \right)  \notag
\end{eqnarray}%
where $V$ is the total volume, the density moments of the bulk free energy
and SGA coefficient are defined as%
\begin{eqnarray}
\overline{\omega }_{n}\left( \rho _{\infty },\rho _{0}\right) &=&\frac{1}{%
\rho _{\infty }-\rho _{0}}\int_{\rho _{0}}^{\rho _{\infty }}\left( \omega
\left( \rho \right) -\omega \left( \rho _{\infty }\right) \right) \left( 
\frac{\rho -\rho _{0}}{\rho _{\infty }-\rho _{0}}\right) ^{n}d\rho \\
\overline{K}_{n}\left( \rho _{\infty },\rho _{0}\right) &=&\frac{1}{\rho
_{\infty }-\rho _{0}}\int_{\rho _{0}}^{\rho _{\infty }}K\left( \rho \right)
\left( \frac{\rho -\rho _{0}}{\rho _{\infty }-\rho _{0}}\right) ^{n}d\rho 
\notag
\end{eqnarray}%
Note that the density arguments have been suppressed (i.e. $\overline{\omega 
}_{0}\equiv \Delta \omega _{0}\left( \rho _{\infty },\rho _{0}\right) $,
etc.), that $\Delta \omega =\omega \left( \rho _{0}\right) -\omega \left(
\rho _{\infty }\right) $, and that the zeroth-order moment $\overline{\omega 
}_{0}\left( \rho _{\infty },\rho _{0}\right) $ is the same as the quantity $%
\overline{\omega }\left( \rho _{\infty },\rho _{0}\right) $ characterizing
the planar interface. Equation (\ref{cluster}) is completely general and
involves no assumptions regarding the size of the cluster.

As shown in Appendix\ref{AppB}, these piecewise-linear  approximations  can be systematically extended to include an arbitrary number of  linear ``links'' in the profile, each with a separate width and slope. It is expected that as the number of links grows, the profile will become closer and closer to the true, free-energy minimizing profile so that this provides an alternative form of solution of the unconstrained problem. As will be noted below, only a few links are needed to obtain high accuracy in the excess free energy. This therefore provides a systematic, and computationally cheap, alternative to the direct solution of the Euler-Lagrange equation for the profile.

\section{Liquid-vapor Nucleation:\ Extending classical nucleation theory}

For a given temperature, there is a unique value of the chemical potential
at which the liquid and vapor phases can coexist. Any other value of the
chemical potential implies stability of one phase over that of the other. In
this Section, the transformation from the metastable phase to the stable
phase is discussed based on the piecewise-linear model for the spherical
cluster. The reason for concentrating on this model, rather than solving the
SGA equations numerically, is to make contact with Classical Nucleation
Theory and to show how it can be extended in a natural way to include
non-classical effects. Furthermore, this simplified version of the theory is
expected to be useful as a starting point for studying more complex systems.

\subsection{Classical Nucleation Theory}

For a given value of $\mu \neq \mu _{coex}$ there is a metastable phase, $%
\rho _{m}$, and a stable phase, $\rho _{s}$ and $\omega \left( \rho
_{m}\right) > \omega \left( \rho _{s}\right) $ while $\omega ^{\prime }\left(
\rho _{m}\right) =\omega ^{\prime }\left( \rho _{s}\right) =0$. The system
is initially in the metastable state $\rho \left( \mathbf{r}\right) =\rho
_{m}$ and homogeneous nucleation proceeds by the spontaneous formation of a
critical cluster. Throughout this discussion, it will be assumed that such
clusters are always spherical. In CNT, the interface between the two phases
is sharp so that the cluster has a well-defined radius, $R$ and, on physical
grounds, the free energy of a cluster is assumed to be given by%
\begin{equation}
\Omega _{CNT}=\frac{4\pi }{3}R^{3}\omega \left( \rho _{0}\right) +\left( V-%
\frac{4\pi }{3}R^{3}\right) \omega \left( \rho _{\infty }\right) +4\pi
R^{2}\gamma _{c}  \label{cntt}
\end{equation}%
where $\gamma _{c}$ is the planar excess free energy per unit area at
coexistence and where the internal and external densities, $\rho_0$ and $\rho_{\infty}$ are to be determined. Extremizing the free energy gives $\omega ^{\prime }\left( \rho
_{0}\right) =\omega ^{\prime }\left( \rho _{\infty }\right) =0$ showing that
the interior and exterior densities are those of the bulk stable and
metastable phases respectively. Then, since $\omega \left( \rho _{s}\right)
<\omega \left( \rho _{m}\right) $ and $\gamma _{c}>0$, the free energy
necessarily has a maximum at the critical radius,%
\begin{equation}
R^{*}=\frac{2\gamma _{c}}{\omega \left( \rho _{m}\right) -\omega \left( \rho
_{s}\right) }
\end{equation}%
giving the free energy barrier%
\begin{equation}
\Omega _{CNT}^{\ast }-\Omega \left( \rho _{m}\right) =\frac{16\pi }{3}\frac{%
\gamma _{c}^{3}}{\left( \omega \left( \rho _{m}\right) -\omega \left( \rho
_{s}\right) \right) ^{2}}.
\end{equation}%
This can then be used to estimate the nucleation rate\cite{Kashchiev} where,
in the course of the analysis, Eq.(\ref{cntt}) is used to determine the
number of non-equilibrium clusters of a given size under the assumption that
the number of clusters of size $N=\frac{4\pi }{3}R^{3}\rho _{0}$ is
proportional to the Boltzmann factor, $\exp \left( -\beta \Omega
_{CNT}\left( N\right) \right) $. Since there is only one variable that can
vary between clusters, i.e. the radius $R$, the implied nucleation pathway
is simply one of increasing radius.

\subsection{Energy-minimized pathways}

Comparison of Eq.(\ref{cntt}) and (\ref{cluster}) shows that the CNT
expression results from the DFT model if one takes the limit $\frac{w}{R}%
\rightarrow 0$ with $\frac{\left( \rho _{\infty }-\rho _{0}\right) ^{2}}{%
2w^{2}}K_{0}$ held fixed which suggests looking at the large $R$ behavior of
the DFT model. Suppose that the width in the planar (i.e. large radius)
limit is $w_{0}$. Then, one can obtain the large-cluster limit by expanding
in the small parameter $\epsilon \equiv w_{0}/R$ so that%
\begin{eqnarray}
w &=&w_{0}+\epsilon w_{1}+... \\
\rho _{0} &=&\rho _{00}+\epsilon \rho _{01}+...  \notag
\end{eqnarray}%
Minimizing the excess free energy with respect to $w$ and $\rho _{0}$ at
fixed $R$ then gives%
\begin{equation}
\Delta \Omega =\frac{4\pi }{3}R^{3}\left( \omega \left( \rho _{00}\right)
-\omega \left( \rho _{\infty }\right) \right) +4\pi R^{2}\gamma \left[
1+\delta \frac{w_{0}}{R}+O\left( \frac{w_{0}}{R}\right) ^{2}\right]
\end{equation}%
with the zeroth and first order densities determined from%
\begin{eqnarray}
\frac{\partial \omega \left( \rho _{00}\right) }{\partial \rho _{00}} &=&0 \\
\rho _{01} &=&-\frac{3}{\rho _{00}-\rho _{\infty }}\frac{\overline{\omega }%
_{0}}{\omega ^{\prime \prime }\left( \rho _{00}\right) }\left( \frac{\Delta
\omega \left( \rho _{00}\right) }{\overline{\omega }_{0}}+\frac{K\left( \rho
_{00}\right) }{\overline{K}_{0}}\right)  \notag
\end{eqnarray}%
where it is understood that $\overline{\omega }_{0}=\overline{\omega }%
_{0}\left( \rho _{\infty },\rho _{00}\right) $, etc. The first equation
specifies that the density in the large $R$ limit is that of the bulk liquid
for the applied chemical potential as expected. The zeroth order width is
found to be 
\begin{equation}
w_{0}=\sqrt{\frac{\left( \rho _{00}-\rho _{\infty }\right) ^{2}}{2\Delta
\omega \left( \rho _{00}\right) }\overline{K}_{0}}
\end{equation}%
and the coefficients for the expansion of the surface tension term are%
\begin{eqnarray}
\gamma &=&2w_{0}\Delta \omega _{0} \\
\delta &=&\frac{\overline{\omega }_{1}}{\overline{\omega }_{0}}+\frac{%
\overline{K}_{1}}{\overline{K}_{0}}+\frac{1}{4}\frac{\rho _{01}}{\rho
_{00}-\rho _{\infty }}\left( \frac{\omega \left( \rho _{00}\right) -\omega
\left( \rho _{\infty }\right) }{\overline{\omega }_{0}}+\frac{K\left( \rho
_{00}\right) }{\overline{K}_{0}}\right) .  \notag
\end{eqnarray}%
Although the radius parameter $R$ is a model-dependent quantity, it can be
related to a more physical quantity, namely the equimolar radius $R_{e}$, via%
\begin{equation}
R_{e}^{3}=\frac{3}{4\pi\left( \rho \left( 0\right) -\rho \left( \infty \right)
\right)}\int \left( \rho \left( \mathbf{r}\right) -\rho \left(
\infty \right) \right) d\mathbf{r}=R^{3}+\frac{1}{4}w\left(
6R^{2}+4Rw+w^{2}\right) .
\end{equation}

While this simple calculation serves to establish a direct link between the
DFT and the CNT free energy model, it leaves open the question of what the
nucleation pathway might be. There is certainly no reason to assume that
clusters grow by increasing the radius parameter in this model while
minimizing the free energy with respect to the other parameters. In fact, as
shown below, there is reason to believe that the radius parameter varies
non-monotonically along the nucleation pathway and that it cannot be used to
parameterize movement along the nucleation pathway:\ i.e. it is not a good
reaction coordinate. Physically, one expects the cluster to grow by adding
atoms so that the excess number of atoms in the cluster should be a useful
reaction coordinate. However, if one minimizes the model free energy with
respect to $R,w$ and $\rho _{0}$ while holding $\Delta N$ constant, and then
again solves perturbatively (now using $\epsilon =\left( \Delta N\right)
^{-1/3}$ as the small parameter) the zeroth order central density is found
to be given by%
\begin{equation}
\frac{\partial \omega \left( \rho _{00}\right) }{\partial \rho _{00}}=\frac{\omega
\left( \rho _{00}\right) -\omega \left( \rho _{\infty }\right)}{\rho_{00}-\rho_{\infty}}
\end{equation}%
which is physically incorrect since it implies that the final phase will not
be the bulk liquid at the applied chemical potential. This happens because
fixing the number of atoms in the cluster has the effect of changing the
chemical potential as can easily be shown by formulating the problem with a
Lagrange multiplier (see Appendix \ref{AppPerturb}).

A final possibility that suggests itself is to minimize at fixed equimolar
radius. This avoids the issue of altering the chemical potential in the
cluster. A perturbative solution, now using $\epsilon =w_{0}/R_{e}$ as the
small parameter, gives 
\begin{equation}
\Delta \Omega =\frac{4\pi }{3}R_{e}^{3}\Delta \omega \left( \rho
_{00}\right) +4\pi R_{e}^{2}\gamma \left[ 1+\delta \frac{w_{0}}{R}+O\left( 
\frac{w_{0}}{R}\right) ^{2}\right]
\end{equation}%
with the zeroth and first order densities%
\begin{eqnarray}
\frac{\partial \omega \left( \rho _{00}\right) }{\partial \rho _{00}} &=&0 \\
\rho _{01} &=&-\frac{3}{\rho _{00}-\rho _{\infty }}\frac{\overline{\omega }%
_{0}}{\omega ^{\prime \prime }\left( \rho _{00}\right) }\left( \left( 1-%
\frac{K\left( \rho _{00}\right) }{\overline{K}_{0}}\right) \frac{\Delta
\omega \left( \rho _{00}\right) }{2\overline{\omega }_{0}}+\frac{K\left(
\rho _{00}\right) }{\overline{K}_{0}\left( \rho _{00}\right) }\right)  \notag
\end{eqnarray}%
a width of 
\begin{equation}
w_{0}=\sqrt{\frac{\left( \rho _{00}-\rho _{\infty }\right) ^{2}}{2\overline{%
\omega }_{0}-\Delta \omega \left( \rho _{00}\right) }K_{0}\left( \rho
_{00}\right) }
\end{equation}%
and the coefficients for the expansion of the surface tension term are%
\begin{eqnarray}
\gamma &=&w_{0}\left( 2\overline{\omega }_{0}-\Delta \omega \left( \rho
_{00}\right) \right) \\
\delta &=&\frac{2\overline{\omega }_{1}-\frac{1}{3}\Delta \omega \left( \rho
_{00}\right) }{2\overline{\omega }_{0}-\Delta \omega \left( \rho
_{00}\right) }+\frac{K_{1}\left( \rho _{00}\right) }{K_{0}\left( \rho
_{00}\right) }-1+\frac{1}{4}\left( \frac{\Delta \omega \left( \rho
_{00}\right) }{2\overline{\omega }_{0}-\Delta \omega \left( \rho
_{00}\right) }+\frac{K\left( \rho _{00}\right) }{K_{0}\left( \rho
_{00}\right) }\right) \frac{\rho _{01}}{\rho _{00}-\rho _{\infty }}%
\allowbreak .  \notag
\end{eqnarray}%
This energy-minimized path therefore gives the correct bulk density and
seems the most likely candidate to be a good approximation to the nucleation
pathway.

\subsection{Transition-state and steepest descent description}

The previous analysis shows that a description of the nucleation pathway in
terms of minimum-energy configurations along some reaction coordinate is not
trivial. Two candidate paths were found, one at fixed parameter $R$ and
another at fixed equimolar radius. This leads to the question as to which
prescription is ''correct''\ or whether there is a less arbitrary means of
constructing such a description.

The process of nucleation of a stable phase from an unstable one is
conceptually similar to that of a chemical reaction or a structural
transition in a finite cluster of molecules. All of these involve the
transition from a higher (free-) energy state to a lower energy state via an
energy barrier. In principle, these processes should be described by
dynamical theories. For the liquid-vapor transition, this would consist of a
hydrodynamic description in which the free energy would enter via the local
pressure. However, dynamical descriptions are computationally expensive and
in the spirit of CNT, it is interesting to ask what can be learned simply
from knowledge of the free energy functional governing the transition. For
example, in CNT the free energy is a function of a single parameter, the
radius, and the transition can be viewed as the growth of the radius from $%
R=0$, the homogeneous metastable phase, to $R\rightarrow \infty $, the
homogeneous stable phase. The goal here is to generalize this picture for
the case in which there are multiple parameters characterizing the
transition, as is the case with the piecewise-linear density profile
discussed above. It should be noted that the same question can be asked with
regard to the SGA free energy functional Eq.(\ref{SGA}), in which case there
is a continuum of parameters, namely the values of $\rho \left( r\right) $
for all points $r$.

The standard approach to the descriptions between (meta-) stable states on
an energy surface, widely used in the examples cited above\cite{Wales},
involves two parts. The first is the determination of \emph{transition states%
} which are defined as saddle points in the free energy surface for which
the Hessian of the (free-)energy function has a single negative eigenvalue.
This is a generalization of the concept of the critical cluster in CNT. The
second element is the determination of \emph{steepest descent pathways}
through the parameter space. These paths start at the transition state and
consist of an initial small movement in the direction of the eigenvector
with the negative eigenvalue. (Actually, there are two paths: one parallel
to the eigenvector and the other anti-parallel.) The steepest descent paths
are then the most efficient paths connecting the transition state to a local
minimum - i.e. to a (meta-)stable state and, one expects, are closely
related to the most likely paths when the processes is driven by thermal
fluctuations. If a model for thermal fluctuations is available, then it is
possible to define instead a most likely path\cite{MaxLikelihoodPath} and of
course, kinetic effects may alter the dynamics substantially. The
steepest-descent paths simply represent the best guess of how the transition
will proceed in the absence of dynamical information.

In the simple case of liquid-vapor nucleation, one anticipates that there
will be a single transition state (the critical cluster) and that the
steepest descent path in one direction will lead to the homogeneous
metastable phase while that in the other will lead to the homogeneous stable
phase. In this case, it is the first part - the path connecting the initial
homogeneous metastable phase to the transition state - which is primarily of
interest.

For the sake of generality, suppose that the free energy depends on $n$
parameters and let $\Gamma $ denote a particular set of those parameters so
that for the piecewise linear model, $n=3$ and $\Gamma =\left( \rho
_{0},R,w\right) $. The transition state is a stationary point so must satisfy%
\begin{equation}
\frac{\partial \Omega }{\partial \Gamma _{a}}=0
\end{equation}%
for all $a=1,...,n$. It can be efficiently located using so-called
eigenvalue following techniques\cite{Eigen}. These are similar to
gradient-following minimization techniques except that one minimizes in
directions corresponding to positive eigenvalues of the Hessian and
maximizes in directions corresponding to negative eigenvalues.

The steepest descent paths are the quickest paths down the gradient starting
at the transition states. The notion of ''quickest'' involves a measure of
distance in parameter space which leads to the question of how to define a
distance between two points $\Gamma _{1}$ and $\Gamma _{2}$ given that the
individual parameters will not in general even have the same units. Since
the picture underlying the model is of a transition from one density profile
to another, and since the concept of distance used to define the steepest
descent directions should be independent of the model, it seems most natural
to use the Euclidean distance in density space,%
\begin{equation}
d^{2}\left[ \rho _{1},\rho _{2}\right] =\int \left( \rho _{1}\left( \mathbf{r%
}\right) -\rho _{2}\left( \mathbf{r}\right) \right) ^{2}d\mathbf{r}
\end{equation}%
which induces a distance measure in parameter space,%
\begin{equation}
d^{2}\left( \Gamma _{1},\Gamma _{2}\right) =\int \left( \rho \left( \mathbf{%
r;\Gamma }_{1}\right) -\rho \left( \mathbf{r;\Gamma }_{2}\right) \right)
^{2}d\mathbf{r}
\end{equation}%
It is clear that in the distance measure in parameter space is not Euclidean
and in fact the induced metric is%
\begin{equation}  \label{metric}
g_{ab}\left( \Gamma \right) =\int \frac{\partial \rho \left( \mathbf{%
r;\Gamma }\right) }{\partial \Gamma _{a}}\frac{\partial \rho \left( \mathbf{%
r;\Gamma }\right) }{\partial \Gamma _{b}}d\mathbf{r.}
\end{equation}%
The steepest descent path for a non-Euclidean geometry is then determined by%
\begin{equation} \label{SDE}
g_{ab}\frac{d\Gamma _{b}}{ds}=\frac{1}{\sqrt{g^{ab}\frac{\partial \Omega }{%
\partial \Gamma _{a}}\frac{\partial \Omega }{\partial \Gamma _{b}}}}\frac{%
\partial \Omega }{\partial \Gamma _{a}}
\end{equation}%
where $s$ is the distance in parameter space\cite{Wales}. As an
illustration, the metric for the piecewise-linear spherical profile is given
explicitly in Appendix \ref{AppMetric}. The procedure is then to find the
transition state which occurs at some point $\Gamma _{0}$, to make a small
displacement of $\Gamma _{0}$ in the direction of the unstable eigenvector
and then to use this as an initial condition for the solution of the
steepest descent equations. Note that the metric, Eq.(\ref{metric}), is only
defined for density profiles which are at least continuous which is one
reason that one cannot use the discontinuous zero-width profile implicitly
assumed in CNT.

Finally, it should be noted that there are alternatives to this procedure.
Techniques such as the Nudged Elastic Band\cite{NEB0,NEB,NEB-CI} and the
String Method\cite{string} are alternative, more heuristic, methods for
determining steepest descent pathways. Both require a measure of distance in
parameter space and so involve the same issues raised here. They are in
general much less computationally demanding than the direct approach
described above and have been used to determine pathways based on the
inhomogeneous density, rather than a parameterization as used here\cite%
{LutskoBubble1,Lutsko_JCP_2008_2}. The reason for using the present approach
is that it seems most in keeping with the spirit of CNT plus the fact that with the simple piecewise-linear models, the direct integration of the steepest descent equations, Eq.(\ref{SDE}), is computationally straightforward.

\section{Comparison of theory and simulation}

To evaluate the SGA theory and the analytic approximations introduced above,
a comparison between the theory and simulations of planar and spherical
liquid-vapor interfaces was carried out for a fluid with the Lennard-Jones
pair potential 
\begin{equation}
v(r) = 4\epsilon\left( \left(\frac{\sigma}{r} \right)^{12}-\left(\frac{\sigma%
}{r} \right)^{6}\right).
\end{equation}%
All forms of the theory require as input the bulk equation of state. In
order to minimize errors arising from this external input, the empirical
Lennard-Jones equation of state of Johnson, Zollweg and Gubbins (JZG)\cite%
{JZG} was used. Nevertheless, it is important to note that the JZG equation
of state is based on simulations covering a limited range of temperatures
and densities with corrections so as to describe the infinite-ranged
Lennard-Jones potential while all of the simulation results used here are
for a finite cutoff. Thus, in all cases, a mean-field correction was added
to the JZG equation of state so as to account for the cutoff\cite{JZG}. It
is therefore expected that the input is most reliable for large cutoffs and
becomes increasingly unreliable as the cutoff is decreased.

\begin{figure*}[tbp]
\includegraphics[angle=-0,scale=0.5]{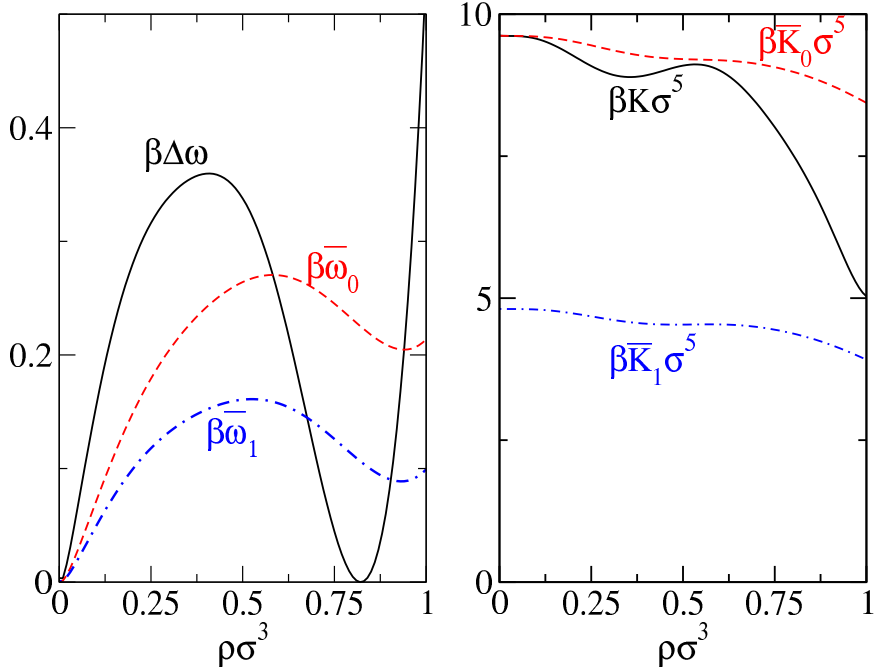}
\caption{The free energy and SGA coefficient and their zeroth and
first-order density moments for a Lennard-Jones fluid with no cutoff at $%
k_{B}T=0.75 \protect\epsilon = 0.58k_{B} T_{c}$ and $\protect\mu = \protect%
\mu_{coex}$.}
\label{fig1}
\end{figure*}

\begin{figure*}[tbp]
\includegraphics[angle=-0,scale=0.5]{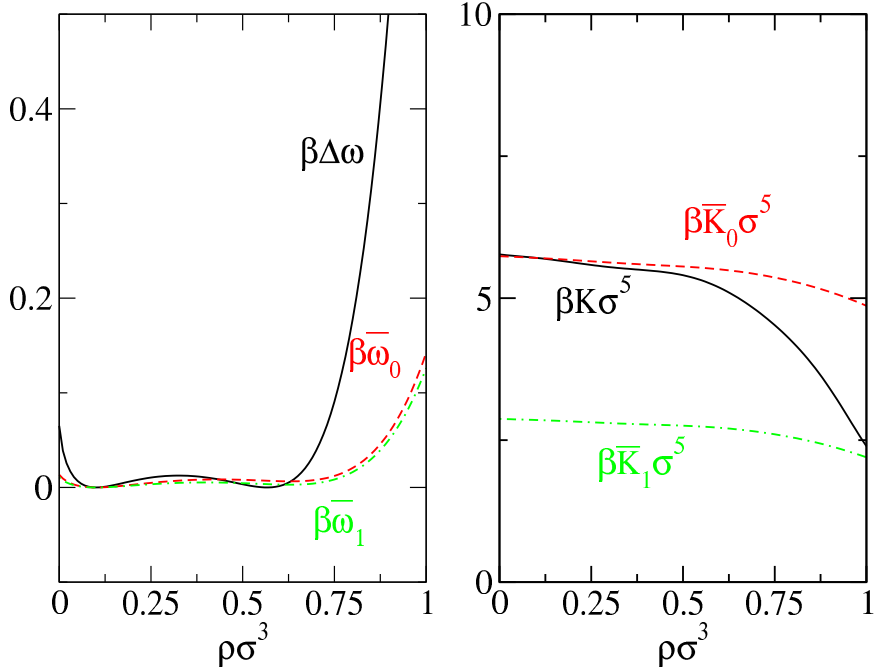}
\caption{The same as Fig.\ref{fig1} but for $k_{B} T=1.2\protect\epsilon %
= 0.92k_{B} T_{c}$.}
\label{fig2}
\end{figure*}

Figure \ref{fig1} shows the free energy, density-averaged free energy and
its first moment, $\Delta \omega \left( \rho ,\rho _{v}\right) ,\bar{\omega}
_{0}\left( \rho ,\rho _{v}\right) $ and $\bar{\omega}_{1}\left( \rho ,\rho
_{v}\right) $, for the Lennard-Jones potential with no cutoff for
coexistence, $\mu =\mu _{coex}$ and at temperature $k_{B}T=0.75\epsilon =
0.58k_{B} T_{c}$ which is just above the triple point (and where $T_c =
1.3\epsilon $ is the critical temperature). Also shown are $K\left( \rho
\right) $, $\bar{K}_{0}\left( \rho ,\rho _{v}\right) $ and $\bar{K}%
_{1}\left( \rho ,\rho _{v}\right) $. The free energy, $\Delta \omega \left(
\rho ,\rho _{v}\right) $, has two minima corresponding to the vapor and the
liquid states separated by a barrier of about $\allowbreak 0.36k_{B}T$. The
density-averaged free energy shows less variation and the first moment is
about half as large. The SGA coefficient varies relatively little as a
function of density thus showing that it is dominated by the
density-independent potential contributions. As a consequence, the
density-averaged value is nearly constant and the first moment is very
nearly half as large, $\bar{K}_{1}\left( \rho ,\rho _{v}\right) \simeq \frac{%
1}{2}\bar{K}_{0}\left( \rho ,\rho _{v}\right) $, over the whole range of
densities. In Fig.\ref{fig2}, the same quantities are shown for $k_{B}T = 1.2\epsilon =
0.92k_{B} T_{c}$. In this case, the difference between the free energy
moments is much less but the SGA coefficient is still dominated by the
density-independent contributions. 

\subsection{Planar interface}

\begin{figure*}[tbp]
\includegraphics[angle=-0,scale=0.5]{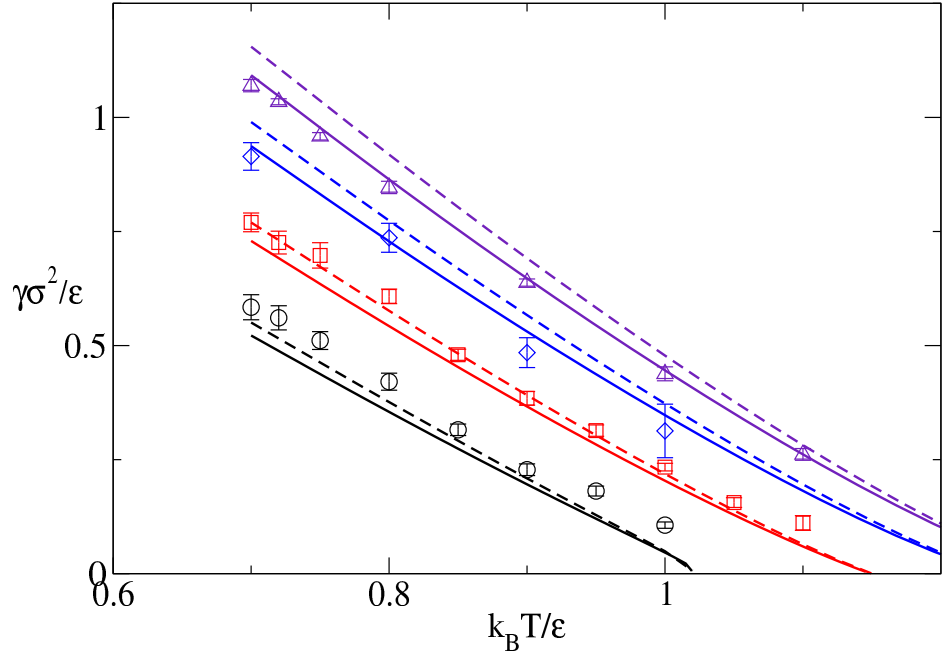}
\caption{The surface tension as a function of temperature for potentials
cutoff at $r_c = 6\protect\sigma$ (upper curves and symbols), $r_c = 4%
\protect\sigma$, $r_c = 3\protect\sigma$ and $r_c = 2.5\protect\sigma$
(lower curves and symbols). The symbols are the simulation data and error
bars reported in Ref.\protect\cite{PGLutsko}, the full lines are from the
SGA calculations and the dashed lines are from the piecewise-linear model.}
\label{fig3}
\end{figure*}

As a first test of the SGA, the profile and excess free energy of the planar
liquid-vapor interface at coexistence was calculated for the Lennard-Jones
potential truncated at various positions, $r_c$, and shifted so that $%
v(r_c)=0$. Figure \ref{fig3} shows the excess free energy per unit area, or
surface tension, as a function of temperature from the calculations as well
as that calculated in the piecewise linear model and values obtained from
Monte Carlo simulations\cite{PGLutsko}. The accuracy of the calculations is
evident. In fact, the accuracy of the SGA appears to rival that of the
underlying DFT (see Ref.\cite{Lutsko_JCP_2008} for comparison). The
piecewise-linear approximation always gives higher values of the surface
tension, as it must since the SGA result is obtained via an unconstrained
minimization of the free energy, but is nevertheless very close to the SGA
result. 

\begin{figure*}[tbp]
\includegraphics[angle=-0,scale=0.5]{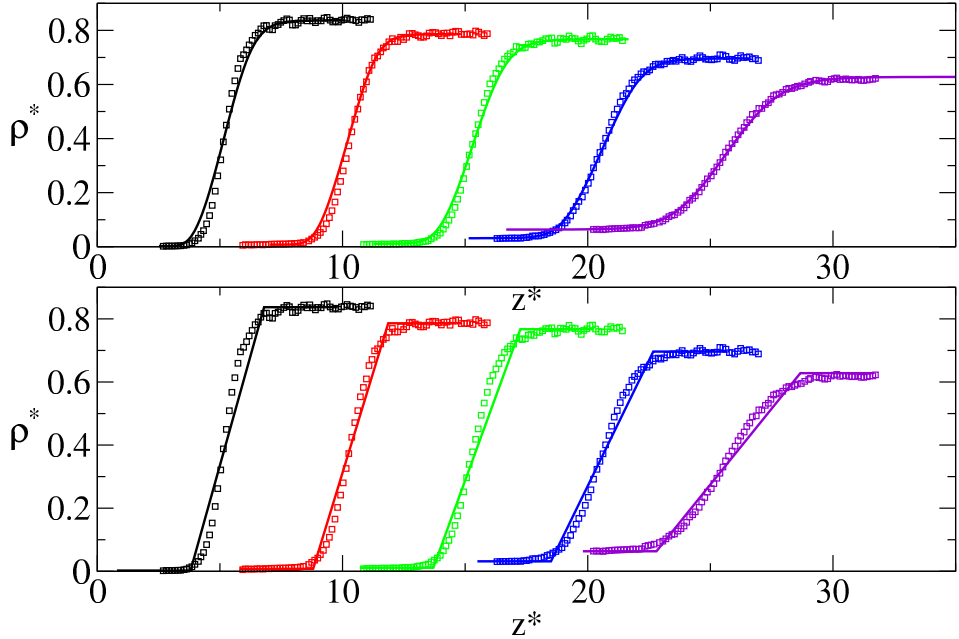}
\caption{(Color online) Density profiles at the liquid-vapor interface
calculated at different temperatures and values of the potential cutoff.
From left to right, the curves correspond to $k_{B}T/\protect\epsilon \equiv
T^*=0.7$ and $r_c^*=5.0$, $T^*=0.7$ and $r_c^*= 2.5$, $T^*=0.8$ and $%
r_c^*=5.0$, $T^*=0.8$ and $r_c^*=2.5$ and $T^*=1.1$ and $r_c^*=5.0$. The
upper panel shows curves calculated in the SGA while the lower one shows the
piece-wise linear approximation. The symbols are the data reported in ref. 
\protect\cite{Mecke-LJ_Interface} and extracted from ref. \protect\cite%
{Katsov} as the original is no longer available\protect\cite{Fisher}. }
\label{fig4}
\end{figure*}

Some calculated profiles are shown in Fig.\ref{fig4} and compared to
simulation data reported in Ref.\cite{PGLutsko}. The SGA profiles are in
reasonable agreement with the simulations although some discrepancy is
apparent, particularly in the narrowest interfaces. The crudeness of the
piecewise-linear approximation is apparent; even so, the widths of the
interfaces are tracked reasonably well as a function of cutoff and
temperature. 

\begin{figure*}[tbp]
\includegraphics[angle=-0,scale=0.5]{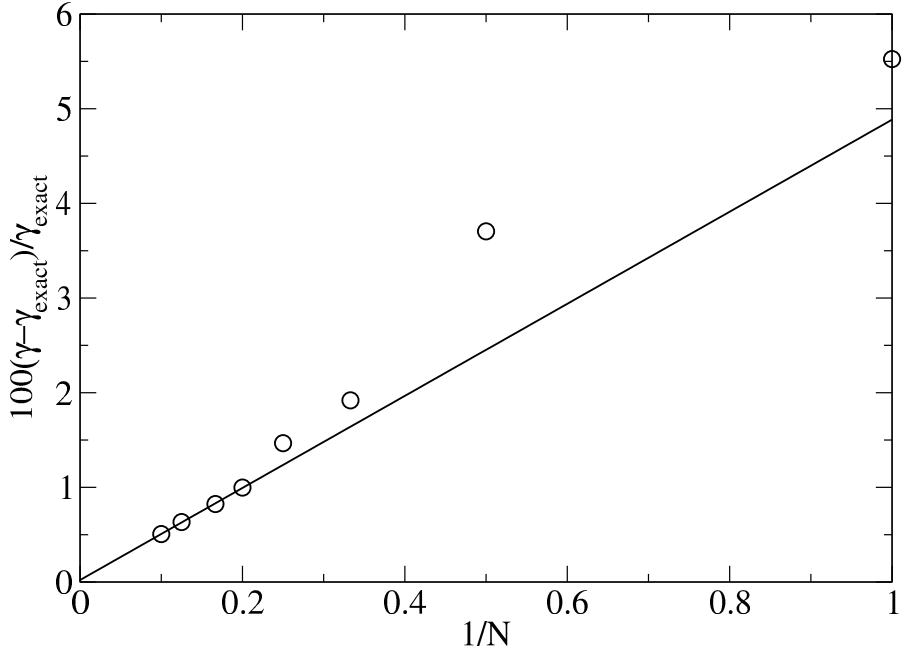}
\caption{The relative error in the surface tension as calculated using piecewise-linear profiles with varying numbers of links, $N$, for $r_{c}^{*}=6$ and for $T^{*}=0.7$. The line shows the linear extrapolation of the profiles with 5,6,8 and 10 links demonstrating convergence to the exact value.}
\label{fig_rel}
\end{figure*}

To illustrate the convergence to the exact result as the number of links in the profile increases, calculations were performed while varying the number of links. The results are shown in Fig.\ref{fig_rel} which shows that the simple single-link profile gives an error of about $5\%$ and that this decreases to about $0.5\%$ for 10 links. Linear extrapolation of the values as a function of $1/N$ gives the exact value. 

\subsection{Clusters}

Small clusters pose more of a challenge since there is no bulk region and
most, if not all, of the atoms in the cluster are affected by the interface.
Furthermore, all clusters are out of equilibrium except the critical cluster
which is in a metastable state. The description of unstable clusters will be
discussed below: here attention is focussed on the transition states: i.e.,
the critical clusters.

Using the SGA, the properties of critical clusters were determined by
solving Eq.(\ref{SGA-1}) in a spherically symmetric geometry using a
relaxation technique\cite{NR}. Given an initial guess of the profile not too
different from the critical cluster, this method relaxes to the critical
cluster automatically. For the analytic model, the eigenvalue-following
technique described above was used. The protocol is the same as described in
Ref.\cite{Lutsko_JCP_2008_2} including a small temperature correction to
account for deficiencies in the equation of state for small cutoffs.

\begin{figure*}[tbp]
\includegraphics[angle=-0,scale=0.5]{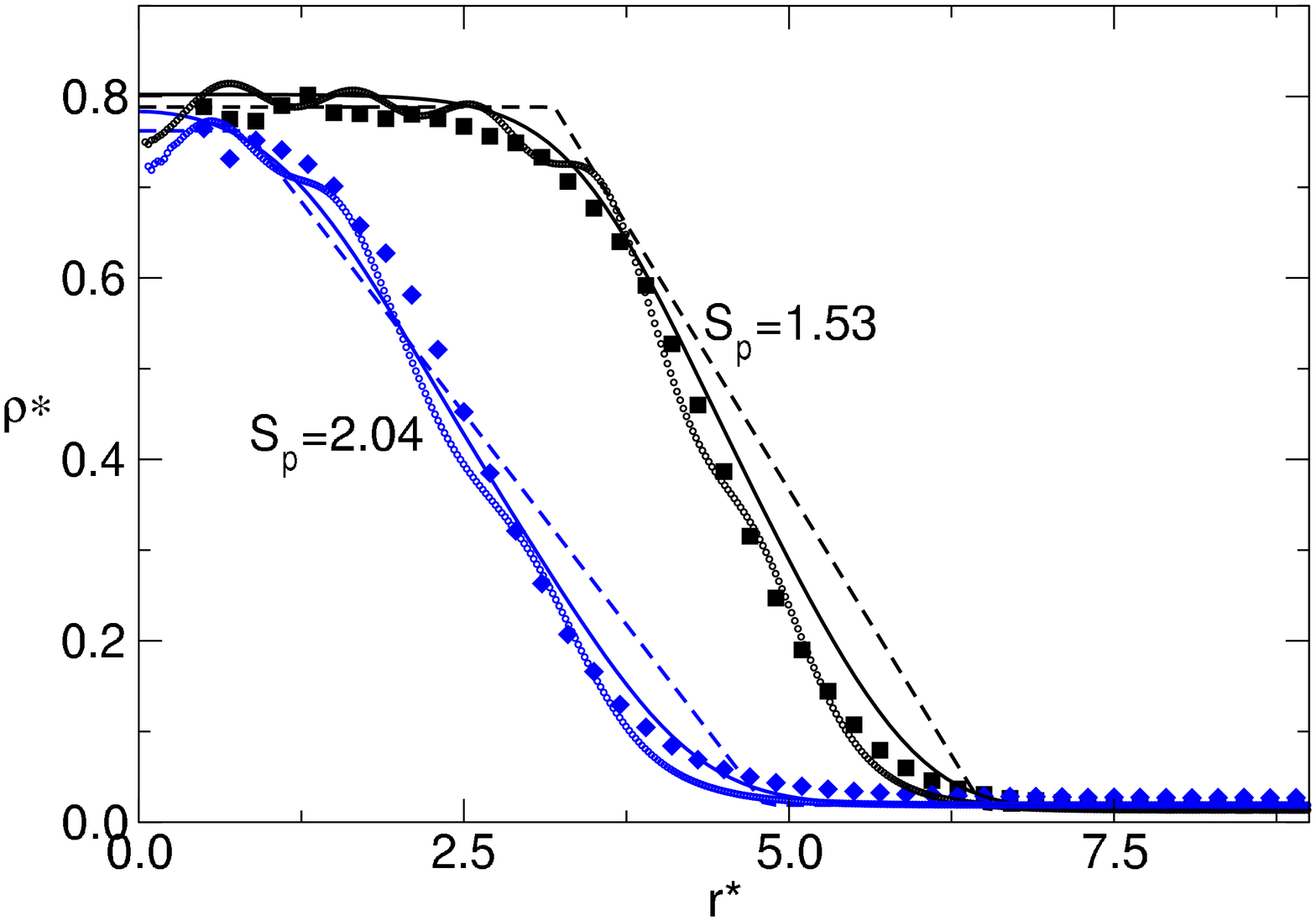}
\caption{(Color online) Density profiles of the critical cluster for two
different values of the supersaturation for the Lennard-Jones potential with cutoff $r_c= 2.5 \sigma$ and $k_{B}T/\epsilon=0.741$. The full symbols are the simulation data
from Ref. \protect\cite{frenkel_gas_liquid_nucleation}, the full lines were
calculated using the SGA, the dashed lines are the piecewise-analytic
approximation and the open circles are the result of the full MC-VDW\cite{Lutsko_JCP_2008_2}.}
\label{fig5}
\end{figure*}

Figure \ref{fig5} shows the density profiles for critical clusters at two
different values of the supersaturation as well as simulation data from ten
Wolde and Frenkel\cite{frenkel_gas_liquid_nucleation}. Note that following
Ref.\cite{frenkel_gas_liquid_nucleation}, supersaturation is defined as the
ratio of the vapor pressure to that at coexistence. The SGA is seen to give
a good description of the critical clusters, although there are greater
differences from simulation than in the case of the planar profiles. In
particular, the SGA profiles have wider interfaces than occurs in the data
and the larger profile, corresponding to lower supersaturation, has somewhat
larger radius than indicated by simulation. Also shown are the analytic
approximations which give a reasonable approximation to the SGA profiles but
which are still wider and therefore compare less well to simulation. To put
these results in context, the underlying MC-VDW DFT model on which the
present SGA is based is in close agreement with the simulated cluster
profiles\cite{Lutsko_JCP_2008_2}.

\begin{figure*}[tbp]
\includegraphics[angle=-0,scale=0.5]{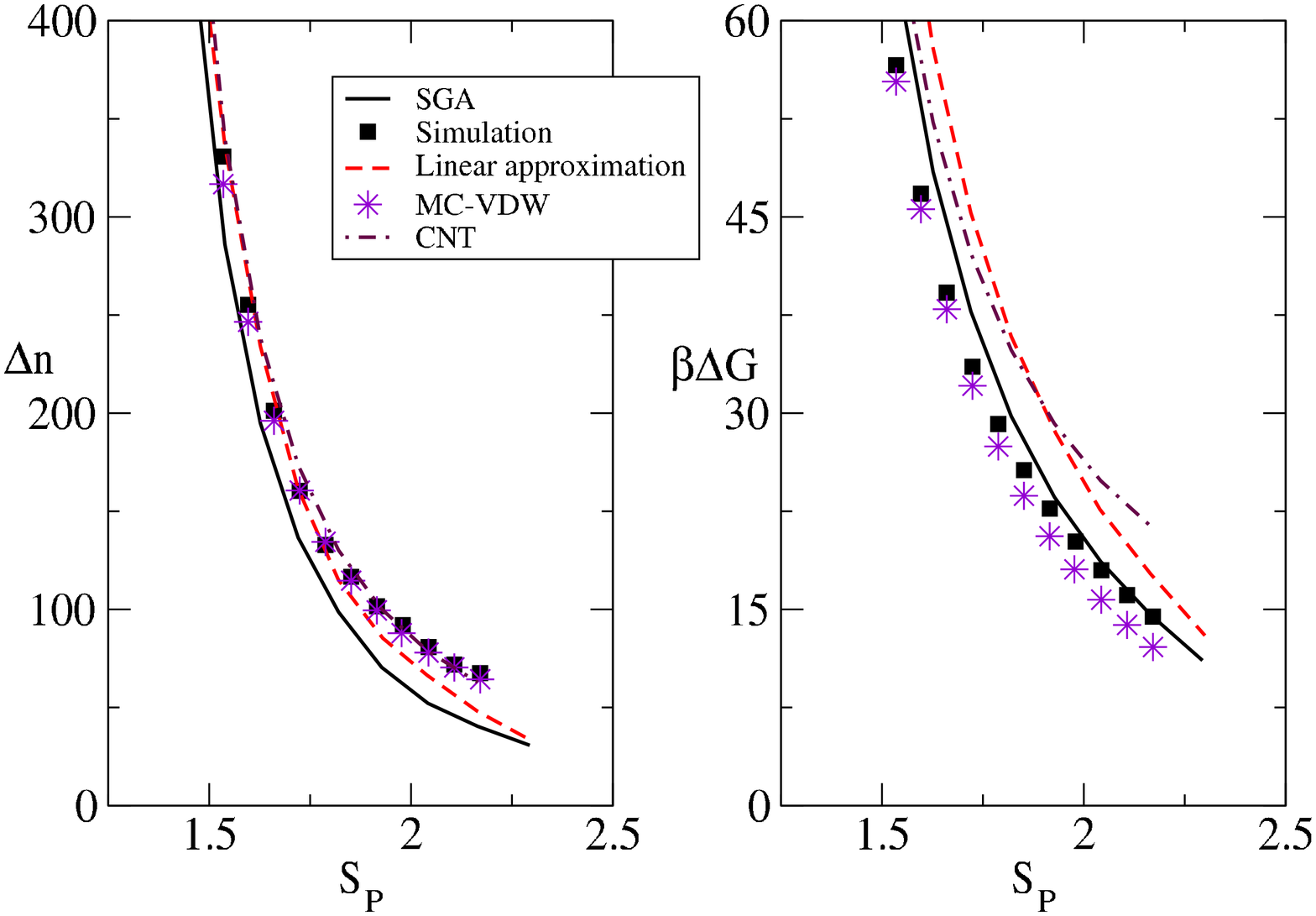}
\caption{(Color online) The properties of the critical cluster as a function
of supersaturation for the Lennard-Jones potential with cutoff $r_c=2.5\sigma$ and $k_{B}T/\epsilon=0.741$. The panel on the left shows the excess number of atoms
in the cluster and that on the right shows the excess free energy. The
squares are the simulation data from Ref. \protect\cite%
{frenkel_gas_liquid_nucleation}, the full lines are the result of solving the SGA, Eq.(\ref{SEL}), the dashed-lines are from the piecewise-linear approximation and, for comparison, the results obtained from the full MC-VDW DFT\cite{Lutsko_JCP_2008_2} are shown as stars and the CNT result is shown as a dashed-dotted line.}
\label{fig6}
\end{figure*}

Figure \ref{fig6} shows the cluster size and excess free energy as a
function of supersaturation as computed from the SGA, the piecewise-linear
profiles, CNT and determined from simulation. The SGA gives a good estimate of
the free energy barrier, but 
systematically underestimates the cluster size due to more rapid convergence to the bulk vapor density. The analytic profiles again
give reasonable approximations to the SGA. As noted by ten Wolde and Frenkel\cite{frenkel_gas_liquid_nucleation} CNT gives good estimates of the cluster size and poorer estimates for the barrier, especially at higher supersaturations.

\subsection{Nucleation Pathways}

\begin{figure*}[tbp]
\includegraphics[angle=-00,scale=0.5]{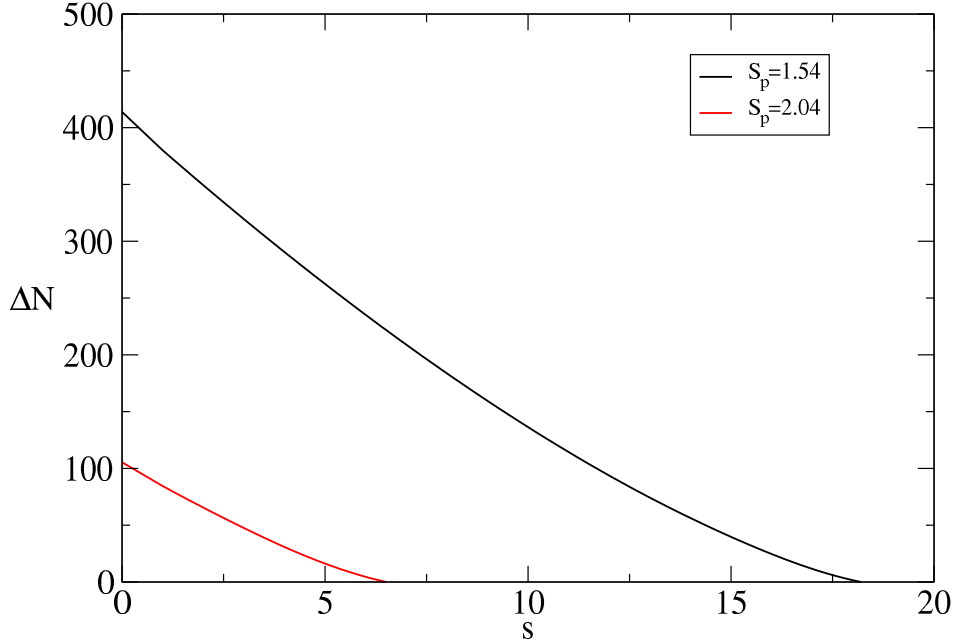}
\caption{(Color online) The excess number of atoms in a cluster as a
function of distance along the steepest-descent paths where $s=0$
corresponds to the critical cluster. The figure shows the results for the
Lennard-Jones potential with a cutoff at $2.5\protect\sigma$ for $T=0.683T_c$
for two different values of the supersaturation and illustrates the fact
that the excess number varies monotonically with distance along the path so
that it can sensibly be used as a reaction coordinate.}
\label{fig7}
\end{figure*}

\begin{figure*}[tbp]
\includegraphics[angle=-0,scale=0.5]{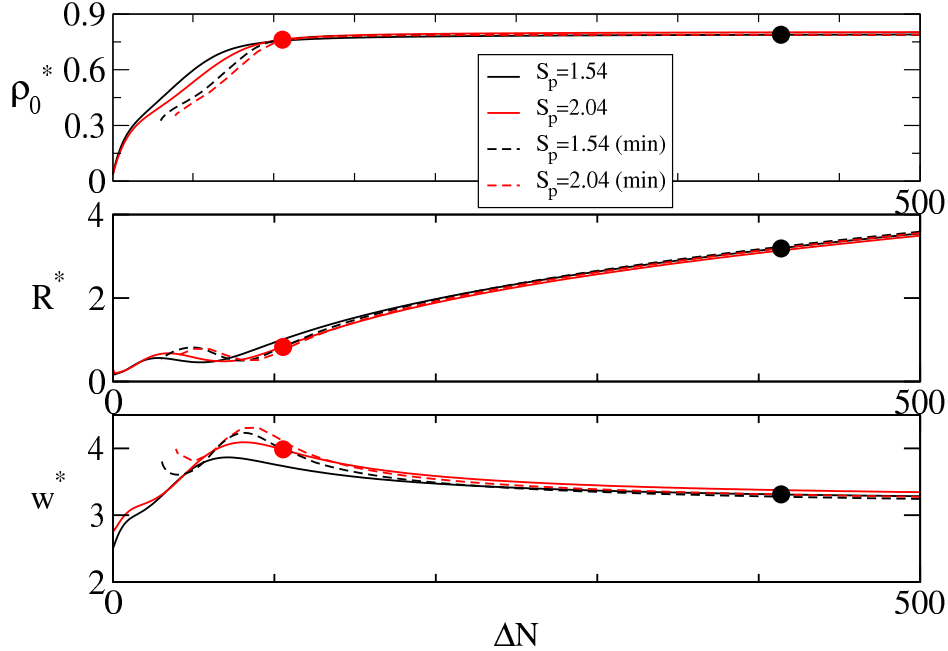}
\caption{(Color online) The inner density, $\protect\rho_0$, the radius, $R$%
, and the width, $w$, as a function of excess number of atoms in a cluster
for the same conditions as described in Fig. \protect\ref{fig7}. The full
curves are the result of solving the steepest descent equations and the
symbols mark the critical clusters. The dashed lines are the minimum-energy
pathways calculated for fixed $\Delta N$.}
\label{fig8}
\end{figure*}

The steepest descent pathways have been calculated by integrating Eq.(\ref{SDE}) for several values of the
supersaturation. Figure \ref{fig7} shows the excess number of atoms in a
cluster as a function of the distance along the pathway for two cases
showing that $\Delta N$ is a monotonic function of the distance and
therefore could serve as a reaction coordinate. Figure \ref{fig8}
illustrates the actual pathways and shows some surprising features. Although
the excess number of atoms is a monotonic function of the path, the size of
the bulk region, parameterized by $R$, and the width of the interface,
parameterized by $w$, are both non-monotonic functions along the steepest-descent
pathway. The width in particular grows with growing droplet size for small
droplets until it reaches a maximum for clusters of about $75$ atoms and
then slowly decreases thereafter. The size of the bulk region shows a local
minimum at about the same point as the width has a maximum and is a
decreasing function of cluster size for small clusters of $30-75$ atoms.

Figure \ref{fig8} also shows the minimum energy pathways obtained by
minimizing with respect to all parameters while holding $\Delta N$ fixed.
The fact, noted above, that the inner density is incorrect for large $\Delta
N$ is evident. However, the critical clusters are accurately determined
since they are stationary points with respect to all parameters. This means
that any energy minimized path, regardless of the constraints, will give the
correct critical cluster. For smaller clusters, the fixed-$\Delta N$ paths
are in qualitative agreement with the steepest descent paths showing the
same non-monotonic behavior of both the width and radius. However, rather
than varying smoothly and continuously as the cluster size goes to zero, it
was found that below a certain cluster size, the energy-minimized path
jumped discontinuously to a solution consisting of a very low central
density, only slightly larger than the gas, and a very large width. This is
not an accident: for the small values of $\Delta N$, there are no
energy-minimized clusters with liquid-like cores. Calculations of the
energy-minimized path at fixed equimolar radius gave good agreement with the
steepest-descent paths for large clusters, but show similar unphysical
behavior (i.e. the absence of liquid-like cores) beginning at larger cluster
sizes than for the fixed $\Delta N$ paths. The conclusion is that while
energy-minimized paths can give qualitative behavior similar to the steepest
descent paths, they do not give a reasonable physical description of the
entire nucleation pathway. This is somewhat at odds with the recent results of Ghosh and Ghosh\cite{Ghosh} who examine energy-minimized paths at fixed radius for sigmoidal profiles and who appear to obtain nontrivial minimizations at all values of the radius. Perhaps this is simply due the use of different equations of state and values of the squared-gradient coefficient or because the piece-wise linear model is too crude. Another possibility  is that it is related to the fact that the sigmoidal profile does not enforce the boundary condition that $d\rho(r)/dr = 0$ at $r=0$ as do the piecewise-linear profiles used here and so are less constrained. Note that this boundary condition is used when solving the SGA Euler-Lagrange equations, Eq.(\ref{SEL}),  for the critical cluster and that without it, derivatives such as $d\rho(x,0,0)/dx$ do not exist at $x=0$. 

\begin{figure*}[tbp]
\includegraphics[angle=-0,scale=0.5]{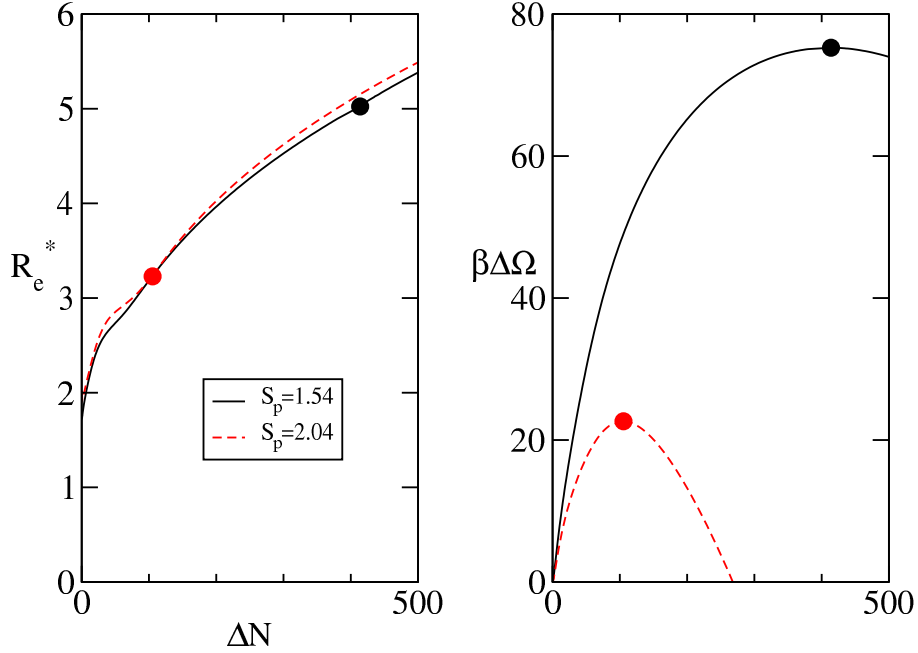}
\caption{(Color online) The equimolar radius (left panel) and the excess
free energy (right panel) as a function of cluster size as calculated along
the steepest descent paths. The calculations were performed for two values
of the supersaturation and the critical cluster in each case is indicated
with a circle.}
\label{fig9}
\end{figure*}

Another interesting feature of the steepest-descent pathways is that the the
interior density is that of the metastable vapor for very small clusters and
increases rapidly as a function of cluster size until the cluster reaches
about $100$ atoms. Since the radius of the bulk region never fully goes to
zero, this gives a very different picture of the formation of small clusters
from that implied in CNT. Recall that in the CNT model, small clusters have
small radii while the central density is always that of the bulk. Here, the
picture is one of increasing density in a bulk region that is always of
finite extent. Figure \ref{fig9} shows the equimolar radius which further
emphasizes this difference. Whereas in the CNT model, the equimolar radius
is equal to the radius of the cluster, and therefore goes to zero for small
clusters, here it is a function of both the radius of the bulk region and
the width and in fact is never small than about $2\sigma $. This is very
similar to results found using the NEB method and a much more sophisticated
DFT model\cite{LutskoBubble1,Lutsko_JCP_2008_2}. Here, however, the physics
behind this behavior is evident. As shown above, the surface free energy is
proportional to the difference in densities inside and outside the droplet
and inversely proportional to the width of the interfacial region. In order
to minimize the free energy for small droplets, which is dominated by the
surface tension contribution, as the size of the droplets decreases, the
difference in densities must decrease as well. Since the free energy is
inversely proportional to the width, the width stabilizes at a finite value
and the $\Delta N\rightarrow 0$ limit is achieved via $\rho _{0}\rightarrow
\rho _{\infty }$ at finite $w$ giving a non-zero equimolar radius.

\section{Conclusions}

The construction of the Squared-Gradient approximation starting with the
Modified-Core van der Waals model for inhomogeneous fluids has been
described. A relatively simple expression for the SGA coefficient was
obtained which requires only the equation of state and interaction potential
as input. The SGA model was shown to give quantitatively accurate surface
free energies and density profiles for planar liquid-vapor interfaces of
Lennard-Jones fluids as a function of temperature and potential cutoff.
Similar comparisons were made for spherical clusters where it was found that
the SGA was less accurate than the full DFT model nevertheless gives
reasonable results.

It was also shown that the SGA could further be approximated using
piecewise-linear density profiles. This model is of course less accurate in
the same examples (planar interfaces and spherical clusters) than the SGA
but is not unreasonable given its simplicity. Aside from providing a rather
simple means to explore the solution of the SGA, the main advantage of the
piecewise-linear model is that it offers a simple bridge between DFT and the
ideas behind Classical Nucleation Theory.

Finally, the use of these tools to construct a description of homogeneous
liquid-vapor nucleation was illustrated. The description was based on an
analogy to chemical and structural transitions and made use of the
transition-state/steepest-descent path methods used in those problems.
Interesting non-classical behavior that was observed included the decrease
of the interior density and the finite equimolar-radius as the size of the
clusters tended to zero. Both effects were linked to the fact that the
effective surface tension is density-dependent (unlike in CNT), a fact that
is immediately evident in the case of the piecewise-linear approximation but
that would be harder to isolate in purely numerical calculations using the
SGA or the original DFT. This density dependence is crucial in following
small clusters as, otherwise, the surface tension would cause $\Delta \Omega$
to diverge as the interfacial width goes to zero. This serves to illustrate
the utility of such simple, yet not unrealistic, approximate methods in
developing physical understanding of the more complex calculations.

\bigskip

\begin{acknowledgments}
I am grateful to Pieter ten Wolde and Daan Frenkel for supplying their
simulation data. This work was supported in part by the European Space
Agency under contract number ESA AO-2004-070.
\end{acknowledgments}

\appendix{}

\section{Evaluation of the coefficients}

\label{AppA}

From Eq.(\ref{bb}), the explicit expression for the core-correction
coefficients is%
\begin{align}
b_{0}& =\frac{3}{\pi d^{3}}\left( \frac{\partial ^{2}}{\partial \overline{%
\rho }^{2}}f_{HS}\left( \overline{\rho }\right) -\frac{\partial ^{2}}{%
\partial \overline{\rho }^{2}}f\left( \overline{\rho }\right) \right)
+3c_{HS}\left( d_{-};\overline{\rho },d\right) +3\beta v\left( d\right) +%
\frac{12}{d^{3}}\int_{d}^{\infty }\beta v\left( r\right) r^{2}dr \\
b_{1}& =-\beta v\left( d\right) -c_{HS}\left( d_{-};\overline{\rho }%
,d\right) -b_{0}  \notag
\end{align}%
The gradient theory coefficient is given by%
\begin{eqnarray}
K &=&\frac{4\pi }{6}\int_{0}^{\infty }c_{HS}\left( r;\overline{\rho }%
;d\right) r^{4}dr+\frac{4\pi }{6}\int_{0}^{d}\left( b_{0}+b_{1}\frac{r}{d}%
\right) r^{4}dr-\frac{4\pi }{6}\int_{d}^{\infty }\beta v\left( r\right)
r^{4}dr \\
&=&\frac{4\pi }{6}\int_{0}^{\infty }c_{HS}\left( r;\overline{\rho };d\right)
r^{4}dr+\frac{\pi d^{5}}{45}\left( 6b_{0}+5b_{1}\right) -\frac{4\pi }{6}%
\int_{d}^{\infty }\beta v\left( r\right) r^{4}dr  \notag \\
&=&\frac{4\pi }{6}\int_{0}^{\infty }c_{HS}\left( r;\overline{\rho };d\right)
r^{4}dr+\frac{\pi d^{5}}{9}\left( -\beta v\left( d\right) -c_{HS}\left(
d_{-};\overline{\rho },d\right) \right) +\frac{\pi d^{5}}{45}b_{0}-\frac{%
4\pi }{6}\int_{d}^{\infty }\beta v\left( r\right) r^{4}dr  \notag \\
&=&\frac{4\pi }{6}\int_{0}^{\infty }c_{HS}\left( r;\overline{\rho };d\right)
r^{4}dr-\frac{2}{45}\pi d^{5}c_{HS}\left( d_{-};\overline{\rho },d\right) +%
\frac{\pi d^{5}}{45}\left( \frac{3}{\pi d^{3}}\left( \frac{\partial ^{2}}{%
\partial \overline{\rho }^{2}}f_{HS}\left( \overline{\rho }\right) -\frac{%
\partial ^{2}}{\partial \overline{\rho }^{2}}f\left( \overline{\rho }\right)
\right) \right)  \notag \\
&&-\frac{2}{45}\pi d^{5}\beta v\left( d\right) +\frac{4\pi }{30}%
\int_{d}^{\infty }\left( 2d^{2}-5r^{2}\right) \beta v\left( r\right) r^{2}dr
\notag
\end{eqnarray}%
Then, using%
\begin{equation}
\frac{\partial ^{2}}{\partial \overline{\rho }^{2}}\left( \beta f_{HS}\left( 
\overline{\rho }\right) -\beta f_{id}\left( \overline{\rho }\right) \right)
=-\int c_{HS}\left( r;\overline{\rho };d\right) d\mathbf{r}
\end{equation}%
gives%
\begin{eqnarray}
K & = & \frac{4\pi }{30}\int_{0}^{\infty }c_{HS}\left( r;\overline{\rho }%
;d\right) \left( 5r^{2}-2d^{2}\right) r^{2}dr-\frac{2}{45}\pi
d^{5}c_{HS}\left( d_{-};\overline{\rho },d\right) -\frac{d^{2}}{15}\frac{%
\partial ^{2}}{\partial \overline{\rho }^{2}}f_{ex}\left( \overline{\rho }%
\right) \\
& - & \frac{2}{45}\pi d^{5}\beta v\left( d\right) + \frac{4\pi }{30}%
\int_{d}^{\infty }\left( 2d^{2}-5r^{2}\right) \beta v\left( r\right) r^{2}dr
\notag
\end{eqnarray}

For both the Percus-Yevick approximation and the more accurate White-Bear
DFT, the hard-sphere DCF can be written as%
\begin{equation}
c_{HS}\left( r;\overline{\rho };d\right) =\left( a_{0}+a_{1}\frac{r}{d}%
+a_{3}\left( \frac{r}{d}\right) ^{3}\right) \Theta \left( d-r\right)
\end{equation}%
giving%
\begin{eqnarray}
\beta K\left( \bar{\rho}\right)  &=&-\frac{\pi d^{5}}{180}a_{3}\left( 
\overline{\rho }d^{3}\right) -\frac{d^{2}}{15}\frac{\partial ^{2}}{\partial 
\overline{\rho }^{2}}\beta f_{ex}\left( \overline{\rho }\right) -\frac{2\pi 
}{45}d^{5}\beta v\left( d\right) +\frac{2\pi }{15}\int_{d}^{\infty }\left(
2d^{2}-5r^{2}\right) \beta v\left( r\right) r^{2}dr \\
&=&-\frac{\pi d^{5}}{180}a_{3}\left( \overline{\rho }d^{3}\right) -\frac{%
d^{2}}{15}\frac{\partial ^{2}}{\partial \overline{\rho }^{2}}\beta
f_{ex}\left( \overline{\rho }\right) +\frac{2\pi }{45}\int_{d}^{\infty
}\left( 3r^{5}-2d^{2}r^{3}\right) \frac{d\beta v\left( r\right) }{dr}dr 
\notag
\end{eqnarray}%

Finally, 
\begin{eqnarray}
\left. \frac{\partial ^{2}}{\partial \overline{\rho }^{2}}f\left( \overline{%
\rho }\right) \right\vert _{V,T} &=&\left. \frac{\partial \mu }{\partial 
\overline{\rho }}\right\vert _{V,T} \\
&=&\left. \frac{\partial \left( \frac{f}{\overline{\rho }}+\frac{\beta P}{%
\overline{\rho }}\right) }{\partial \overline{\rho }}\right\vert _{V,T} 
\notag \\
&=&\overline{\rho }^{-1}\left. \frac{\partial f}{\partial \overline{\rho }}%
\right\vert _{V,T}-\frac{f}{\overline{\rho }^{2}}+\overline{\rho }%
^{-1}\left. \frac{\beta P}{\partial \overline{\rho }}\right\vert _{V,T}-%
\frac{\beta P}{\overline{\rho }^{2}}  \notag \\
&=&\overline{\rho }^{-1}\left. \frac{\beta P}{\partial \overline{\rho }}%
\right\vert _{V,T}  \notag \\
&=&\overline{\rho }^{-2}\kappa _{T}  \notag
\end{eqnarray}%
where the isothermal compressibility is%
\begin{equation}
\kappa _{T}=\beta _{T}^{-1}=-V\frac{\partial \beta P}{\partial V}
\end{equation}%
So%
\begin{equation}
K=-\frac{\pi d^{5}}{180}a_{3}-\frac{d^{2}}{15}\overline{\rho }^{-2}\kappa
_{T}-\frac{2\pi }{45}d^{5}\beta v\left( d\right) + \frac{2\pi }{45}\int_{d}^{\infty
}\left( 3r^{5}-2d^{2}r^{3}\right) \frac{d\beta v\left( r\right) }{dr}dr 
\end{equation}

For packing fraction $\eta =\frac{\pi }{6}\rho d^{3}$, the Percus-Yevick
approximation gives%
\begin{equation}
a_{3}=-\frac{\eta }{2}\frac{\left( 1+2\eta \right) ^{2}}{\left( 1-\eta
\right) ^{4}}
\end{equation}%
while in the White-Bear approximation,%
\begin{equation}
a_{3}=\frac{-3+10\eta -15\eta ^{2}+5\eta ^{3}}{\left( 1-\eta \right) ^{4}}-%
\frac{3\ln \left( 1-\eta \right) }{\eta }
\end{equation}%
\bigskip

\section{More general planar interfaces}

\label{AppB}

The piecewise-linear model is easily extended to include an arbitrary number
of pieces. The density profile becomes%
\begin{equation*}
\rho \left( z\right) =\left\{ 
\begin{array}{c}
\rho _{-\infty },\;z<0 \\ 
\rho _{-\infty }+\left( \rho _{1}-\rho _{-\infty }\right) \frac{z}{w_{1}}%
,\;0<z<w_{1} \\ 
\rho _{1}+\left( \rho _{2}-\rho _{1}\right) \frac{z-w_{1}}{w_{2}}%
,\;w_{1}<z<w_{1}+w_{2} \\ 
\rho _{2}+\left( \rho _{3}-\rho _{2}\right) \frac{z-w_{1}-w_{2}}{w_{3}}%
,\;w_{1}+w_{2}<z<w_{1}+w_{2}+w_{3} \\ 
... \\ 
\rho _{\infty },\;w_{1}+w_{2}+...+w_{n}<z%
\end{array}%
\right.
\end{equation*}%
Defining 
\begin{equation*}
z_{i}=\sum_{j=1}^{i}w_{j}
\end{equation*}%
this can be written as%
\begin{equation}
\rho \left( z\right) =\left\{ 
\begin{array}{c}
\rho _{-\infty },\;z<0 \\ 
\rho _{i-1}+\left( \rho _{i}-\rho _{i-1}\right) \frac{z-z_{i-1}}{w_{i}}%
,\;z_{i-1}<z<z_{i} \\ 
\rho _{\infty },\;z_{n}<z%
\end{array}%
\right.
\end{equation}%
with the identifications $\rho _{0}=\rho _{-\infty }$, $\rho _{n}=\rho
_{\infty }$. The free parameters are $\rho _{i}$ for $1\leq i\leq n-1$ and $%
w_{i}$ for $1\leq i\leq n$ giving a total of $2n-1$ parameters. Then, the
excess energy is 
\begin{eqnarray}
\gamma &=&\frac{\Omega -\Omega _{coex}}{V}=\sum_{i=1}^{n}%
\int_{z_{i-1}}^{z_{i-1}+w_{i}}\left\{ \omega \left( \rho \left( z\right)
\right) -\omega _{coex}+\frac{1}{2}K\left( \rho \left( z\right) \right)
\left( \frac{\rho _{i}-\rho _{i-1}}{w_{i}}\right) ^{2}\right\} dz \\
&=&\sum_{i=1}^{n}\left\{ w_{i}\int_{\rho _{i-1}}^{\rho _{i}}\left( \omega
\left( x\right) -\omega _{coex}\right) dx+\frac{\left( \rho _{i}-\rho
_{i-1}\right) ^{2}}{2w_{i}}\int_{\rho _{i-1}}^{\rho _{i}}K\left( x\right)
dx\right\}  \notag
\end{eqnarray}%
which is simply the sum of the contribution of each link in the profile.

\section{Perturbative Solution for the spherical profile at constant
particle number}

\label{AppPerturb}

The free energy is%
\begin{equation}
\Delta \Omega =\frac{4\pi }{3}R^{3}\Delta \omega +4\pi R^{2}w\left( 
\overline{\omega }_{0}+2\overline{\omega }_{1}\left( \frac{w}{R}\right) +%
\overline{\omega }_{2}\left( \frac{w}{R}\right) ^{2}+\frac{\left( \rho
_{\infty }-\rho _{0}\right) ^{2}}{2w^{2}}\left( \overline{K}_{0}+2\overline{K%
}_{1}\left( \frac{w}{R}\right) +\overline{K}_{2}\left( \frac{w}{R}\right)
^{2}\right) \right)
\end{equation}%
which is to be minimized at constant particle number,%
\begin{equation}
N=\int \left( \rho \left( \mathbf{r}\right) -\rho _{\infty }\right) d\mathbf{%
r}=\frac{\pi }{3}\left( \rho _{0}-\rho _{\infty }\right) \left( 2R+w\right)
\left( 2Rw+2R^{2}+w^{2}\right)
\end{equation}%
Introducing a Lagrange multiplier,$\lambda $, and setting%
\begin{equation*}
0=\frac{\partial }{\partial \Gamma }\left( \Delta \Omega \left( R,\rho
_{0},w\right) -\lambda \left( N-\frac{\pi }{3}\left( \rho _{0}-\rho _{\infty
}\right) \left( 2R+w\right) \left( 2Rw+2R^{2}+w^{2}\right) \right) \right)
\end{equation*}%
for $\Gamma =R,\rho _{0},w$ and $\lambda $ gives, after some simplification,%
\begin{eqnarray}
0 &=&\Delta \omega +2\frac{w}{R}\left( \overline{\omega }_{0}+\overline{%
\omega }_{1}\left( \frac{w}{R}\right) +\frac{\left( \rho _{0}-\rho _{\infty
}\right) ^{2}}{2w^{2}}\left( \overline{K}_{0}+\overline{K}_{1}\left( \frac{w%
}{R}\right) \right) \right) \\
&&+\lambda \frac{1}{3}\left( \rho _{0}-\rho _{\infty }\right) \left(
3+3\left( \frac{w}{R}\right) +\left( \frac{w}{R}\right) ^{2}\right)  \notag
\\
0 &=&\frac{1}{3}\frac{\partial \Delta \omega }{\partial \rho _{0}}+\frac{w}{R%
}\left( \frac{\partial \overline{\omega }_{0}}{\partial \rho _{0}}+2\frac{%
\partial \overline{\omega }_{1}}{\partial \rho _{0}}\left( \frac{w}{R}%
\right) +\frac{\partial \overline{\omega }_{2}}{\partial \rho _{0}}\left( 
\frac{w}{R}\right) ^{2}\right)  \notag \\
&&+\frac{w}{R}\frac{\left( \rho _{0}-\rho _{\infty }\right) ^{2}}{2w^{2}}%
\left( \frac{\partial \overline{K}_{0}}{\partial \rho _{0}}+2\frac{\partial 
\overline{K}_{1}}{\partial \rho _{0}}\left( \frac{w}{R}\right) +\frac{%
\partial \overline{K}_{2}}{\partial \rho _{0}}\left( \frac{w}{R}\right)
^{2}\right)  \notag \\
&&+\frac{w}{R}\frac{\left( \rho _{0}-\rho _{\infty }\right) }{w^{2}}\left( 
\overline{K}_{0}+2\overline{K}_{1}\left( \frac{w}{R}\right) +\overline{K}%
_{2}\left( \frac{w}{R}\right) ^{2}\right) +\lambda \frac{1}{12}\left( 4+6%
\frac{w}{R}+4\left( \frac{w}{R}\right) ^{2}+\left( \frac{w}{R}\right)
^{3}\right)  \notag \\
0 &=&\left( \overline{\omega }_{0}+4\overline{\omega }_{1}\left( \frac{w}{R}%
\right) +3\overline{\omega }_{2}\left( \frac{w}{R}\right) ^{2}+\frac{\left(
\rho _{0}-\rho _{\infty }\right) ^{2}}{2w^{2}}\left( -\overline{K}_{0}+%
\overline{K}_{2}\left( \frac{w}{R}\right) ^{2}\right) \right)  \notag \\
&&+\lambda \frac{1}{12}\left( \rho _{0}-\rho _{\infty }\right) \left(
6+8\left( \frac{w}{R}\right) +3\left( \frac{w}{R}\right) ^{2}\right)  \notag
\\
0 &=&N-\frac{\pi }{3}\left( \rho _{0}-\rho _{\infty }\right) R^{3}\left(
4+6\left( \frac{w}{R}\right) +4\left( \frac{w}{R}\right) ^{2}+\left( \frac{w%
}{R}\right) ^{3}\right)  \notag
\end{eqnarray}%
These are solved perturbatively using $\epsilon \equiv N^{-1/3}$ as a small
parameter where the expansion of the various quantities is assumed to take
the form%
\begin{eqnarray}
R &=&\epsilon ^{-1}R_{0}+R_{1}+\epsilon R_{2}+... \\
\rho _{0} &=&\rho _{00}+\epsilon \rho _{01}+...  \notag \\
w &=&w_{0}+\epsilon w_{1}+...  \notag \\
\lambda &=&\lambda _{0}+\epsilon \lambda _{1}+...  \notag
\end{eqnarray}%
Then, the lowest order equations are 
\begin{eqnarray}
0 &=&\omega \left( \rho _{00}\right) -\omega \left( \rho _{\infty }\right)
+\lambda _{0}\left( \rho _{00}-\rho _{\infty }\right) \\
0 &=&\frac{\partial \omega \left( \rho _{00}\right) }{\partial \rho _{00}}%
+\lambda _{0}  \notag \\
0 &=&\overline{\omega }_{0}\left( \rho _{\infty },\rho _{00}\right) -\frac{%
\left( \rho _{\infty }-\rho _{00}\right) ^{2}}{2w_{0}^{2}}\overline{K}%
_{0}\left( \rho _{\infty },\rho _{00}\right) +\lambda _{0}\frac{1}{2}\left(
\rho _{00}-\rho _{\infty }\right)  \notag \\
0 &=&1-\frac{4\pi }{3}\left( \rho _{00}-\rho _{\infty }\right) R_{0}^{3} 
\notag
\end{eqnarray}%
giving%
\begin{eqnarray}
\lambda _{0} &=&-\frac{\omega \left( \rho _{00}\right) -\omega \left( \rho
_{\infty }\right) }{\rho _{00}-\rho _{\infty }} \\
\frac{\partial \omega \left( \rho _{00}\right) }{\partial \rho _{00}} &=&%
\frac{\omega \left( \rho _{00}\right) -\omega \left( \rho _{\infty }\right) 
}{\rho _{00}-\rho _{\infty }}  \notag \\
w_{0}^{2} &=&\frac{\left( \rho _{\infty }-\rho _{00}\right) ^{2}}{2\overline{%
\omega }_{0}\left( \rho _{\infty },\rho _{00}\right) -\left( \omega \left(
\rho _{00}\right) -\omega \left( \rho _{\infty }\right) \right) }\overline{K}%
_{0}  \notag \\
R_{0}^{3} &=&\frac{3}{4\pi \left( \rho _{00}-\rho _{\infty }\right) }  \notag
\end{eqnarray}%
Notice that the second equation can be written as%
\begin{equation}
\frac{\partial f\left( \rho _{00}\right) }{\partial \rho _{00}}=\mu -\lambda
_{0}
\end{equation}%
thus showing the shift of the chemical potential arising from the constraint.

\bigskip

\section{Metric}

\label{AppMetric}

The metric is calculated using Eq.(\ref{metric}). The result for a piecewise-linear profile with a single link is%
\begin{eqnarray}
g_{\rho \rho } &=&\frac{2\pi }{15}\left(
10R^{3}+10R^{2}w+5Rw^{2}+w^{3}\right) \\
g_{\rho R} &=&\left( \rho _{0}-\rho _{\infty }\right) \frac{\pi }{3}\left(
6R^{2}+4Rw+w^{2}\right)  \notag \\
g_{\rho w} &=&\left( \rho _{0}-\rho _{\infty }\right) \frac{\pi }{15}\left(
10R^{2}+10Rw+3w^{2}\right)  \notag \\
g_{RR} &=&\left( \rho _{0}-\rho _{\infty }\right) ^{2}\frac{4\pi }{3}\frac{%
3R^{2}+3Rw+w^{2}}{w}  \notag \\
g_{Rw} &=&\left( \rho _{0}-\rho _{\infty }\right) ^{2}\frac{\pi }{3}\frac{%
6R^{2}+8Rw+3w^{2}}{w}  \notag \\
g_{ww} &=&\left( \rho _{0}-\rho _{\infty }\right) ^{2}\frac{2\pi }{15}\frac{%
10R^{2}+15Rw+6w^{2}}{w}  \notag
\end{eqnarray}


%

\end{document}